\documentclass[conference]{IEEEtran}
\ifCLASSINFOpdf
\else
\fi
\usepackage{url}
\usepackage{amsmath}
\usepackage{algpseudocode}
\usepackage{algorithm}
\usepackage{hhline}
\usepackage{graphicx}
\graphicspath{ {figures/} }
\usepackage{svg}
 \usepackage{booktabs}
 \usepackage{balance}
 \usepackage{url}
 \usepackage{hyperref}

\begin{document}
%
\title{\textbf{AIR-FI:} Generating Covert Wi-Fi Signals from Air-Gapped Computers}

\author{\IEEEauthorblockN{Mordechai Guri}
\IEEEauthorblockA{Ben-Gurion University of the Negev, Israel\\Cyber-Security Research Center\\
gurim@post.bgu.ac.il\\ demo video: \url{https://www.youtube.com/watch?v=vhNnc0ln63c} \\ air-gap research page:  \url{http://www.covertchannels.com} }}
 


%


\maketitle

\begin{abstract}
In this paper, we show that attackers can exfiltrate data from air-gapped computers via Wi-Fi signals. Malware in a compromised air-gapped computer can generate signals in the Wi-Fi frequency bands. The signals are generated through the memory buses - no special hardware is required. Sensitive data can be modulated and secretly exfiltrated on top of the signals. We show that nearby Wi-Fi capable devices (e.g., smartphones, laptops, IoT devices) can intercept these signals, decode them, and send them to the attacker over the Internet. To extract the signals, we utilize the physical layer information exposed by the Wi-Fi chips. We implement the transmitter and receiver and discuss design considerations and implementation details. We evaluate this covert channel in terms of bandwidth and distance and present a set of countermeasures. Our evaluation shows that data can be exfiltrated from air-gapped computers to nearby Wi-Fi receivers located a distance of several meters away.

\end{abstract}


%
\IEEEpeerreviewmaketitle

\section{Introduction}
One of the initial phases in the kill chain of advanced persistent threats (APTs) is  infiltrating the network of the target organization. To achieve this goal, adversaries may use attack vectors, such phishing emails, compromised websites, malicious documents, exploit kits, and other types of online attacks \cite{bahrami2019cyber}.      
 
\subsection{Isolated, Air-Gapped Networks}    
When highly sensitive or confidential information is involved, an organization may resort to air-gapped networks. Such networks are disconnected from the Internet logically and physically, with any type of wired or wireless connection to the Internet strictly prohibited \cite{May2018G94:online}. Certain sectors may maintain their data within air-gapped networks, including financial, defense, and critical infrastructure sectors. In many cases, operational technology (OT) networks are also kept isolated from the Internet to protect the physical processes and machinery used to carry them out \cite{PostDrea48:online}. Classified networks such as the Joint Worldwide Intelligence Communications System are also known to be air-gapped \cite{Storefro90:online}.

\subsection{Infecting Air-Gapped Networks}    
Despite the high degree of isolation, air-gapped networks are not immune to cyber attacks. To penetrate highly secure networks, motivated adversaries may employ complex attack vectors, such as sabotaging the supply chain, compromising a third-party software, using malicious insiders, and exploiting deceived insiders \cite{case2016analysis}\cite{PostDrea48:online}. These techniques allow the attackers to insert targeted malware into systems within the isolated environment.
 
One of the most famous incidents in which the air-gap was breached involved the Stuxnet worm  which targeted supervisory control and data acquisition (SCADA) systems and destroyed an estimated 1,000 centrifuges at an Iranian uranium enrichment facility \cite{kushner2013real}. In 2018, the US Department of Homeland Security accused Russian hackers of penetrating the internal network of America's electric utilities  \cite{Nobigdea65:online}. In 2019, the media reported that the Kudankulam Nuclear Power Plant was the target of a successful cyber attack earlier that year  \cite{AnIndian12:online}. In addition, sophisticated malware, such as SymonLoader \cite{Tickes43:online} and other advanced persistent threats capable of compromising air-gapped networks, were found in the wild \cite{AFannyEq68:online,CycldekB23:online,bahrami2019cyber}.

\subsection{Air-Gap Exfiltration}
Once the attacker has taken his/her its initial step into the air-gapped network, he/she moves on to the next phases of the kill chain. In these subsequent phases sensitive data is collected, including: documents, files, keylogging, credentials, and biometric information. In the case of Internet connected networks the data is exfiltrated through covert channels within Internet protocols (e.g., HTTPS, FTP, SSH, and SMTP \cite{zander2007survey}). However, in isolated air-gapped networks, the attacker must use unconventional communication techniques to leak the data out - methods which are referred to as air-gap covert channels \cite{Guri:2018:BAM:3200906.3177230}. Over the years, various types of air-gap covert channels have been introduced. For example, malware may exploit electromagnetic radiation from various computer components to transmit data \cite{guri2014airhopper,kuhn1998soft,kuhn2002compromising,vuagnoux2009compromising,guri2015gsmem}. Acoustic \cite{carrara2014acoustic,guri2020fansmitter}, optical \cite{loughry2002information,Guri2017,Guri2017a}, thermal \cite{Guri2015a}, magnetic \cite{GURI2021115}, and electric \cite{guri2019powerhammer} air-gap covert channels have also been demonstrated over the past 20 years \cite{Carrara2016}.

\subsection{Our Contribution}
In this paper we introduce a new type of covert channel that exploits Wi-Fi to leak data from air-gapped networks. The AIR-FI attack introduced in this paper does not require Wi-Fi  related hardware in the air-gapped computers. Instead, we show that an attacker can exploit the DDR SDRAM buses to generate electromagnetic emissions in the 2.4 GHz Wi-Fi bands and encode binary data on top of it. We also show that nearby Wi-Fi receivers, such smartphones, laptops, and Internet of Things (IoT) devices, can receive and decode the modulated data, and then send it to the attacker via the Internet.

The AIR-FI covert channel has the following characteristics:

\begin{itemize}
	\item {\textbf{Requires no Wi-Fi transmitter.}} The method doesn't require any type of Wi-Fi hardware in the air-gapped computer. Instead, it uses the computer memory hardware (DDR SDRAM)  to generate the signals. 
	
	\item {\textbf{Requires no special privileges.}} The transmitting code does not require special privileges (e.g., root), kernel drivers, or access to hardware resources. Furthermore, it can be initiated from an ordinary user space process.
	
	\item {\textbf{Works in virtual machines (VMs).}} The covert channel works effectively, even from within an isolated virtual machine.
	
	\item {\textbf{Has many potential receivers.}} Modern IT environments are equipped with many types of Wi-Fi capable devices: smartphones, laptops, IoT devices, sensors, embedded systems and smart watches and other wearables devices. The attacker can potentially hack such equipment to receive the AIR-FI transmissions from air-gapped computers. 
\end{itemize}

The rest of this paper is organized as follows: Related work is presented in Section \ref{sec:related}. The attack model is discussed in Section \ref{sec:attack}. Technical background on DDR SDRAM and Wi-Fi is provided in Section \ref{sec:tech}. Sections \ref{sec:trans} and \ref{sec:rec}, respectively, contain details on signal generation and modulation, and data transmission and reception. In Section \ref{sec:eval} we present the evaluation and measurement results. A set of countermeasures is discussed in Section \ref{sec:counter}, and we conclude in Section \ref{sec:conclusion}.

\section{Related Work}
\label{sec:related}
Air-gap covert channels are classified into seven main categories: electromagnetic, magnetic, electric, acoustic, thermal, optical and vibrational. 

Kuhn showed that it is possible to exploit the electromagnetic emissions from the computer display unit to conceal data \cite{kuhn1998soft}. AirHopper, presented in 2014, is a new exfiltration malware, capable of leaking data from air-gapped computers to a nearby smartphone via FM radio waves emitted from the screen cable \cite{guri2014airhopper,guri2017bridging}. In 2015, Guri et al presented GSMem \cite{guri2015gsmem}, malware that transmit data from air-gapped computers to nearby mobile-phones using cellular frequencies. USBee is malware that uses the USB data buses to generate electromagnetic signals \cite{guri2016usbee}. 

In order to prevent electromagnetic leakage, Faraday cages can be used to shield sensitive systems. Guri et al presented ODINI \cite{guri2019odini} and MAGNETO \cite{GURI2021115}, two types of malware that can exfiltrate data from Faraday-caged air-gapped computers via magnetic fields generated by the computer's CPU. With MAGNETO  the authors used the magnetic sensor integrated in smartphones to receive covert signals. 

In 2019, researchers show how to leak data from air-gapped computers by modulating binary information on the power lines \cite{guri2019powerhammer}. The data is modulated and conducted to the power lines and received by an adversary tapping the wires.

Several studies have proposed the use of optical emanations from computers for covert communication. Loughry introduced the use of keyboard LEDs \cite{loughry2002information}. Guri used the hard drive indicator LED \cite{Guri2017}, USB keyboard LEDs \cite{guri2019ctrl}, router and switch LEDs \cite{guri2018xled}, and security cameras and their IR LEDs \cite{guri2019air}, in order to exfiltrate data from air-gapped computers. Data can also be leaked optically through fast blinking images or low contrast bitmaps projected on the LCD screen \cite{guri2016optical}.

Hanspach \cite{hanspach2014covert} used inaudible sound to establish a covert channel between air-gapped laptops equipped with speakers and microphones. Guri et al introduced Fansmitter \cite{guri2020fansmitter}, Diskfiltration \cite{guri2017acoustic}, and CD-LEAK \cite{guri2020cd} malware which facilitates the exfiltration of data from an air-gapped computer via noise intentionally generated from the PC fans, hard disk drives \cite{guri2017acoustic}, and CD/DVD drives \cite{guri2020cd}. In these methods, the transmitting computer does not need to be equipped with audio hardware or an internal or external speaker. Researchers also showed that the computer fans generate vibrations which can be sensed by a nearby smartphone using the accelerometer sensor \cite{guri2020air}. Other papers presented malware that covertly turns the speakers and earphones connected to a PC into a pair of eavesdropping microphones when a standard microphone is muted, turned off, or not present \cite{guri17speake} \cite{guri2018mosquito}. Recently, researchers demonstrated how malware can turn the computer power supply into out-of-band speaker in order to exfiltrate information \cite{guri2020power}.  

Guri et al introduced BitWhisper \cite{guri2015bitwhisper} and HOTSPOT \cite{guri2019hotspot}, thermal-based covert channels enabling bidirectional communication between air-gapped computers by hiding data in temperature changes. The heat which is generated by the CPU can be received by temperature sensors of computers or smartphones, decoded, and sent to the attacker.

\begin{table}[]
	\centering
	\caption{Summary of existing air-gap covert channels}
	\label{tab:related}
	\begin{tabular}{|l|l|}
		\hline
		Type            & Method                                                                                                                                                                                                                                         \\ \hline
		Electromagnetic & \begin{tabular}[c]{@{}l@{}}AirHopper (FM radio) \cite{guri2014airhopper,guri2017bridging}\\ GSMem (cellular frequencies) \cite{guri2015gsmem}\\ USBee (USB bus emission) \cite{guri2016usbee} \\ AIR-FI (Wi-Fi frequencies)\end{tabular}                                                                              \\ \hline
		Magnetic        & \begin{tabular}[c]{@{}l@{}}MAGNETO (CPU-generated magnetic fields) \cite{GURI2021115} \\ ODINI (Faraday shield bypass)  \cite{guri2019odini}  \end{tabular}                                                                                                                               \\ \hline
		Electric        & PowerHammer (power lines) \cite{guri2019powerhammer}   \\ \hline
		Acoustic        & \begin{tabular}[c]{@{}l@{}}Fansmitter (computer fan noise) \cite{guri2020fansmitter}\\ DiskFiltration (hard disk noise) \cite{guri2017acoustic}\\ Ultrasound \cite{hanspach2014covert} \\ MOSQUITO (speaker-to-speaker) \cite{guri17speake} \cite{guri2018mosquito}\\ POWER-SUPPLAY (Play sound from Power-Supply) \cite{guri2020power} \\ CD-LEAK (sound from CD/DVD drives) \cite{guri2020cd} \end{tabular} \\ \hline
		Thermal         & \begin{tabular}[c]{@{}l@{}}BitWhisper (CPU generated heat) \cite{guri2015bitwhisper}\\ HOTSPOT (
			CPU generated heat received by a smartphone) \cite{guri2019hotspot} \end{tabular}                                                                                                                                \\ \hline
		Optical         & \begin{tabular}[c]{@{}l@{}}LED-it-GO (hard drive LED) \cite{Guri2017}\\ VisiSploit (invisible pixels) \cite{guri2016optical} \\ Keyboard LEDs  \cite{loughry2002information} \cite{guri2019ctrl} \\ Router LEDs \cite{guri2018xled}\\ aIR-Jumper (security cameras and infrared) \cite{guri2019air}\end{tabular}                                                                \\ \hline
		Vibrations      & AiR-ViBeR (computer fan vibrations) \cite{guri2020air}                                                                                                                                                                                                                    \\ \hline
	\end{tabular}
\end{table}
Table \ref{tab:related} summarizes the existing air-gap covert channels.

\section{Attack Model}
\label{sec:attack}

\subsection{Infecting the Air-Gapped Network}
In a preliminary stage, the air-gapped network is infected with an APT. In a typical APT kill chain, the attackers research their targets and carefully plan the attacks \cite{bahrami2019cyber}. After defining the initial target, attackers might install malware on the network via various infection vectors: supply chain attacks, contaminated USB drives, social engineering techniques, or by using malicious insiders or deceived employees. Note that infecting air-gapped networks can be accomplished, as demonstrated by the attacks involving Stuxnet \cite {langner2011stuxnet}, Agent.Btz \cite{grant2009cyber}, and other malware \cite{TheEpicT20:online,RedOcto50:online,AFannyEq68:online}. At that point, the APT might exploit vulnerabilities to spread in the network in order to strengthen its foothold. 

\subsection{Infecting Wi-Fi Devices}
The attacker must infect Wi-Fi capable devices in the area of the air-gapped network.  Such devices might be smartphones of visitors or employees, desktop and laptop computers with wireless networking, or IoT devices with Wi-Fi transceivers.
Since the devices use wireless networking they can be infected through Wi-Fi. Compromising the devices can be done by exploiting vulnerabilities in the Wi-Fi hardware/software or via flaws in the network protocols. Such attacks were demonstrated on smartphones \cite{vanhoef2014advanced}, laptops with Wi-Fi network interface cards (NICs) \cite{berghel2005wifi}, and a wide range of IoT devices such as smart bulbs \cite{ronen2016extended}, smart locks \cite{ho2016smart}, and more \cite{stute2019billion,nawir2016internet}.     

The compromised Wi-Fi capable devices are installed with the receiver side of the malware. In most cases, the malicious code will be executed within the kernel driver or the firmware which drive the Wi-Fi hardware. The malware collects the Wi-Fi signals, detects the covert AIR-FI transmission, decodes the information, and sends it to the attacker over the Internet.    

\subsection{Data Exfiltration}
\begin{figure}  
	\centering
	\includegraphics[width=1\linewidth]{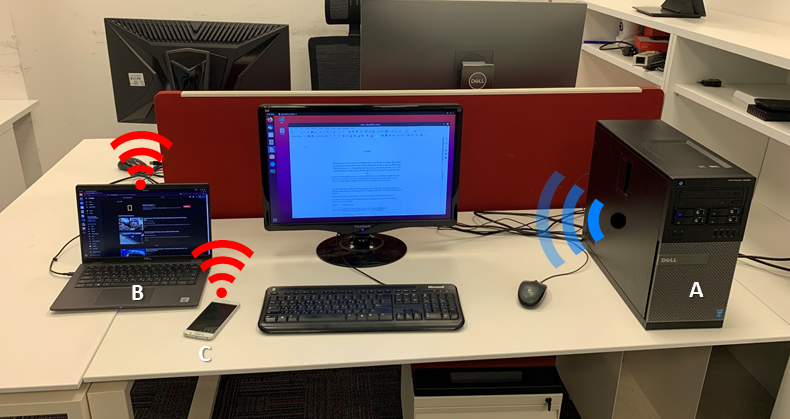}
	\caption{illustration of the AIR-FI attack. Malware in the air-gapped computer (A) uses the  DDR memory to generate signals in the 2.4 GHz Wi-Fi frequency band. Binary information is modulated on top of the signals and received by a nearby Wi-Fi receivers (e.g., laptop (B) and smartphone (C)).}
	\label{fig:attackmodel}
\end{figure}

As a part of the exfiltration phase, the attacker might collect data from the compromised computers. The data can be documents, key logging, credentials, encryption keys, etc. Once the data is collected, the malware initiates the AIR-FI covert channel. It encodes the data and transmits it to the air (in the Wi-Fi band at 2.4 GHz) using the electromagnetic emissions generated from the DDR SDRAM buses. The attack is illustrated in Figure \ref{fig:attackmodel}. Malware in the air-gapped computer (A) uses the  memory to generate signals in the 2.4 GHz Wi-Fi frequency band. Binary information is modulated on top of the signals and received by a nearby Wi-Fi receivers (e.g., laptop (B) and smartphone (C)).

%

\section{Technical Background}
\label {sec:tech}
\subsection{DDR SDRAM}
The double data rate (DDR) synchronous dynamic random-access memory (SDRAM) is the type of memory modules integrated into modern motherboards. The DDR technology doubles the bus bandwidth by transferring data on both the rising and falling edges of the memory bus clock. In DDR SDRAM the bus bandwidth is referred to in megabits per second. The bandwidth $B$ is calculated by the formula $B = (f*2*l)/8$, where $f$ is the memory bus clock rate and $l$ is the width of the line transfer. Another important parameter of memory modules is the Column Address Strobe (CAS) latency, also known as the CL. This is the time delay between when the read command is delivered to the memory and the beginning of the data response.

\subsection{DDR Memory Bus}
\begin{figure}  
	\centering
	\includegraphics[width=0.8\linewidth]{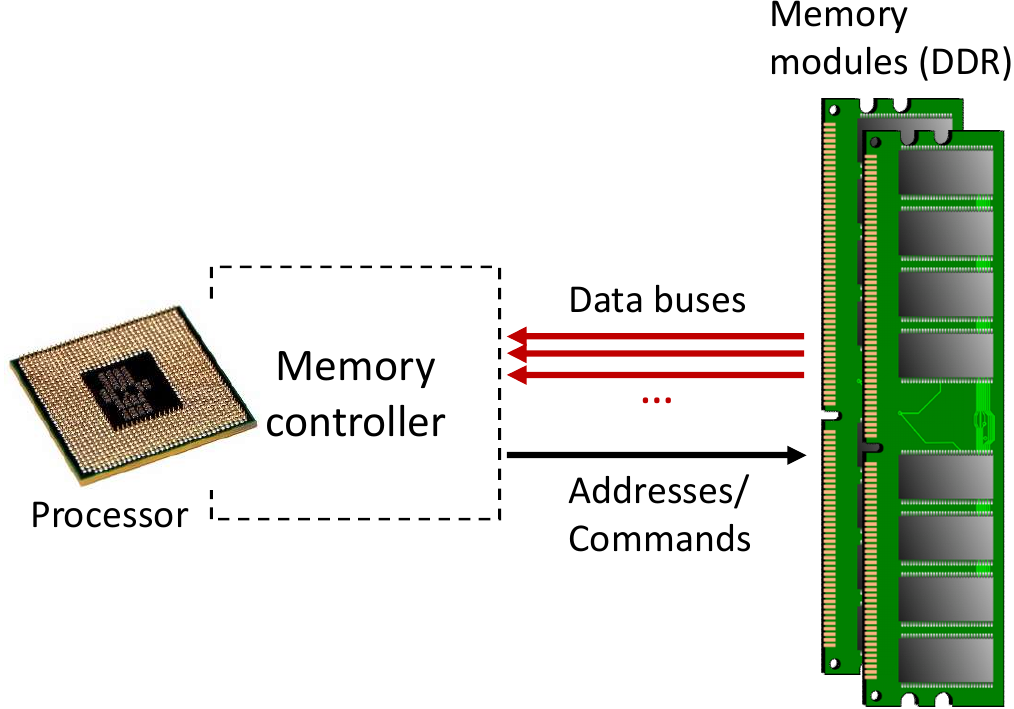}
	\caption{DDR SDRAM memory buses.}
	\label{fig:cont}
\end{figure}

Data is exchanged between the CPU and the memory over dedicated buses (Figure \ref{fig:cont}). The memory buses maintain two types of signals: (1) the address bus which transfers addresses and commands, and (2) the data bus (DQ bus) which transfers the actual data. The address bus sends commands and instructions from the controller to the SDRAM. The bus is synchronized to the clock (CLK) signals, with the signals on the address bus being sampled by the SDRAMs on the rising edge of the CLK signal.

The memory buses generate electromagnetic radiation at a frequency correlated to its clock frequency and harmonics. For example, DDR4-2400 emits electromagnetic radiation at around 2400 MHz. 


\subsection{Overclocking/Underclocking}
The memory modules provide the BIOS/UEFI (Unified Extensible Firmware Interface) a set of frequencies that it can operate at. This information is defined according to the JEDEC (Joint Electron Device Engineering Council) specification, and it is passed during the boot through a mechanism called Serial Presence Detect (SPD). Intel allows the standard timing parameters of the installed memory to be changed via a specification called Extreme Memory Profile (XMP). With XMP the user can modify the parameters of the memory such as the frequency and CAS latency. Changing the operating frequency of the memory modules is referred to as overclocking (for increasing the frequency) and underclocking/downclocking (for decreasing the frequency). 

\subsection{Wi-Fi Frequency Bands} 
The IEEE 802.11 standard defines the frequency ranges in the electromagnetic spectrum allowed for Wi-Fi communications. There are several versions of the 802.11 standard. These standards define factors, such as the frequency ranges, bandwidths and distances. Today, most Wi-Fi chips support the 802.11b/g/n standards. 
The 802.11b/g/n standards are often referred to as the 2.4 GHz band. A range of 2.400 - 2.490 GHz is the most widely used and certified range available for Wi-Fi. The standards define a total of 14 channels in the 2.4 GHz band, but only 11 of these channels are allowed in all countries. The first 11 channels have a space of 5 MHz between them, and there is a space of 12 MHz between channel 13 and 14. A common bandwidth of a Wi-Fi channel is 20 MHz which means that signals of adjacent channels may interfere with each other. Table \ref{tab:freqlist} contains a list of the regulated Wi-Fi channels supported by the 802.11b/g/n standards. 

\begin{table}[]
	\centering
	\caption{List of the regulated Wi-Fi channels (802.11b/g/n)}
	\label{tab:freqlist}
	\begin{tabular}{@{}llllll@{}}
		\toprule
		Channel & Center (MHz) & \begin{tabular}[c]{@{}l@{}}Range\\ (MHz)\end{tabular} & \begin{tabular}[c]{@{}l@{}}North\\ America\end{tabular} & Japan    & Others \\ \midrule
		1       & 2412     & 2401-2423                                                         & Yes                                                     & Yes      & Yes             \\
		2       & 2417     & 2406-2428                                                         & Yes                                                     & Yes      & Yes             \\
		3       & 2422     & 2411-2433                                                         & Yes                                                     & Yes      & Yes             \\
		4       & 2427     & 2416-2438                                                         & Yes                                                     & Yes      & Yes             \\
		5       & 2432     & 2421-2443                                                         & Yes                                                     & Yes      & Yes             \\
		6       & 2437     & 2426-2448                                                         & Yes                                                     & Yes      & Yes             \\
		7       & 2442     & 2431-2453                                                         & Yes                                                     & Yes      & Yes             \\
		8       & 2447     & 2436-2458                                                         & Yes                                                     & Yes      & Yes             \\
		9       & 2452     & 2441-2463                                                         & Yes                                                     & Yes      & Yes             \\
		10      & 2457     & 2446-2468                                                         & Yes                                                     & Yes      & Yes             \\
		11      & 2462     & 2451-2473                                                         & Yes                                                     & Yes      & Yes             \\
		12      & 2467     & 2456-2478                                                         & Canada only                                             & Yes      & Yes             \\
		13      & 2472     & 2461-2483                                                         & No                                                      & Yes      & Yes             \\
		14      & 2484     & 2473-2495                                                         & No                                                      & 11b only & No              \\ \bottomrule
	\end{tabular}
\end{table}

\section{Transmission}
\label{sec:trans}
In this section we present the signal generation technique, data modulation, and data transmission protocol. 

\subsection{Electromagnetic Emission}
There are two types of electromagnetic emissions that emanate from memory buses.

\begin{itemize}
	\item {\textbf{Persistent Emission.}} An electromagnetic emission continuously generated by the memory controller regardless of the activity in the address/data buses. This radiation spans the entire spectrum of the DDR SDRAM frequency when the computer is turned on. 
	
	\item {\textbf{Triggered Emission.}} An electromagnetic emission generated from the electronic activities (current flow) in the data bus. This emission is correlated to the memory read/write operations executed by processes currently running in the system. 
\end{itemize}

\subsection{Signal Generation}
Based on the above observations, we used two techniques to generate Wi-Fi signals from the an air-gapped computer.

\begin{itemize}
	\item {Memory operations.} We transfer data in the data bus to generate an electromagnetic emission at the frequency of the memory modules. Since the clock speed of memory modules is typically around the frequency of 2.4 GHz or its harmonics, the memory operations generate electromagnetic emissions around the IEEE 802.11b/g/n Wi-Fi frequency bands.  
	
	\item {Memory operations + clocking.} When the operational frequency of the memory modules is not near the 2.4 GHz frequency or its harmonics, we initially overclock/downclock the memory speed to the frequency of Wi-Fi bands or its harmonics. The overlocking/downclocking operation can be done programmatically or at the BIOS/UEFI configuration level. Following the frequency adjustments, we perform the memory operation schemes described above to generate emissions at the Wi-Fi frequency band. Note that malware which are capable of reconfiguring BIOS/UEFI were found in the wild \cite{UEFIasam17:online,TrickBot45:online}. 
\end{itemize}
\subsection{Channel Interference}
\begin{figure}  
	\centering
	\includegraphics[width=0.8\linewidth]{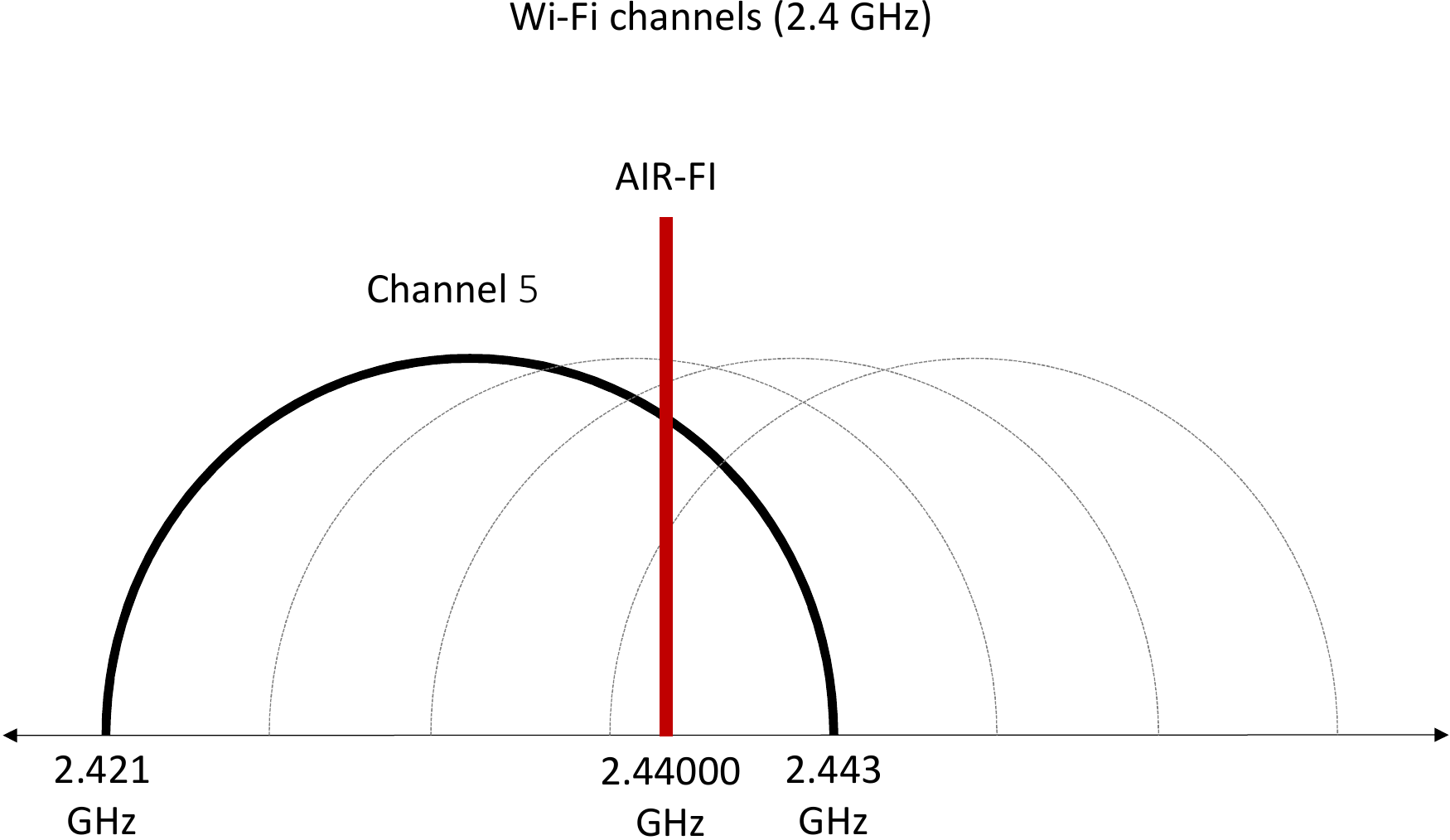}
	\caption{AIR-FI channel interference.}
	\label{fig:channels}
\end{figure}

The generated emission from the data bus interfere with the Wi-Fi channels. The interferences in the corresponding channel can be measured at the PHY layer of the 802.11 protocol stack. The operation is illustrated in figure \ref{fig:channels}. In this case the AIR-FI signals are generated at 2.44000 GHz. The signal are interfering with channels 5-8.

\subsection{Modulation}
Algorithm \ref{algmod} shows the signal modulation process using the memory operation technique using on-off keying (OOK) modulation.  
The \texttt{modulateRAM} function receives the array of bits to transmit (\texttt{bits}) and the bit time in milliseconds (\texttt{bitTimeMillis}). This function iterates over the bits and according to the current bit, the algorithm determines the operation to perform during a bit time period. If the bit is '1' (line 4) it performs a series of memory write operations which consists of sequential memory copying between two arrays each the size of 1 MB size each (lines 6-7). This loop effectively generates the emission from the data bus. If the bit is '0' the algorithm sleeps for a bit time period, which stops the emission from the RAM bus.
\begin{algorithm}
	\caption{modulateRAM (bits, bitTimeMillis)} 
	\label{algmod} 
	\begin{algorithmic}[1] 
		
		\State $ bitEndTime \gets getCurrentTimeMillis() $
		
		\For{$ bit\ in\ bits $}
		\State $ bitEndTime \gets bitEndTime + bitTimeMillis $
		\If {$ bit == 1 $} 
		\While{$getCurrentTimeMillis() <  bitEndTime$}
		\State $ memcopy(array1,array2) $ 
		\State $ memcopy(array2,array1) $ 
		\EndWhile
		\Else
		\State $ sleep(bitTimeMillis) $ 
		\EndIf
		
		\EndFor  
		
	\end{algorithmic}
\end{algorithm}

\subsubsection{Multi cores}
\begin{figure}  
	\centering
	\includegraphics[width=0.8\linewidth]{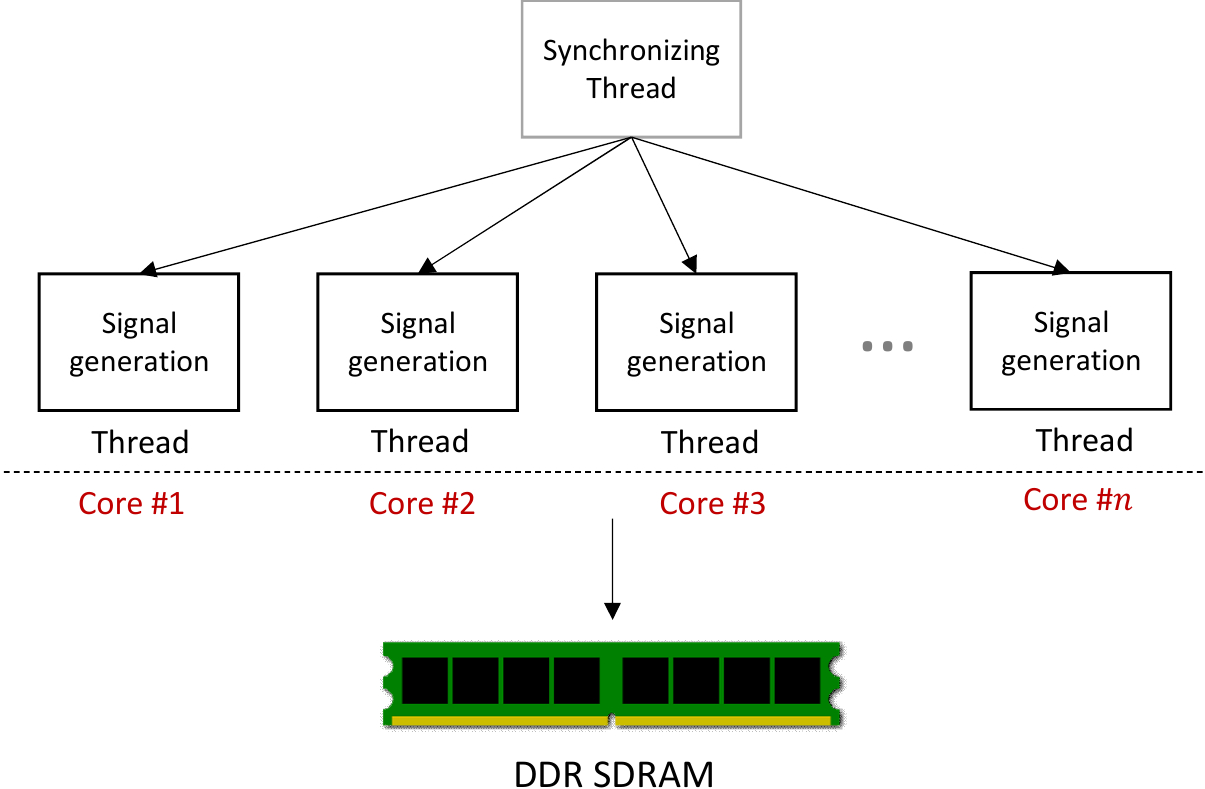}
	\caption{Signal generation with concurrent threads.}
	\label{fig:coressynch}
\end{figure}

The signal generation algorithm shown above runs on a single CPU core. In order to amplify the signal, we execute the code generation in several concurrent threads, where each thread is bound to a specific core. The memory operation of the threads are synchronized by a governor thread using the POSIX thread functions, such as  \texttt{thread\_barrier\_wait}. Signal generation with concurrent threads is depicted in Figure \ref{fig:coressynch}.  

\subsection{Packets}
The data is transmitted in packets that consist of a preamble, payload and error-detecting code. 

\begin{itemize}
	\item {Preamble.} The packet begins with a \texttt{0xAA} hex value. This sequence of \texttt{10101010} in binary allows the receiver to synchronize with the beginning of each packet and determine the carrier amplitude and one/zero thresholds. 
	
	\item {Payload.} The payload is the raw binary data transmitted within the packet. It consists of 32 bits.  
	
	\item {Error detection.} For error detection, we use the CRC-8 (a cyclic redundancy check) error detection algorithm. The CRC is calculated on the payload data and added at the end of each packet. On the receiver side, if the received CRC and the calculated CRC differ, the packet is omitted. 
\end{itemize}

\begin{figure}
	\centering
	\includegraphics[width=\linewidth]{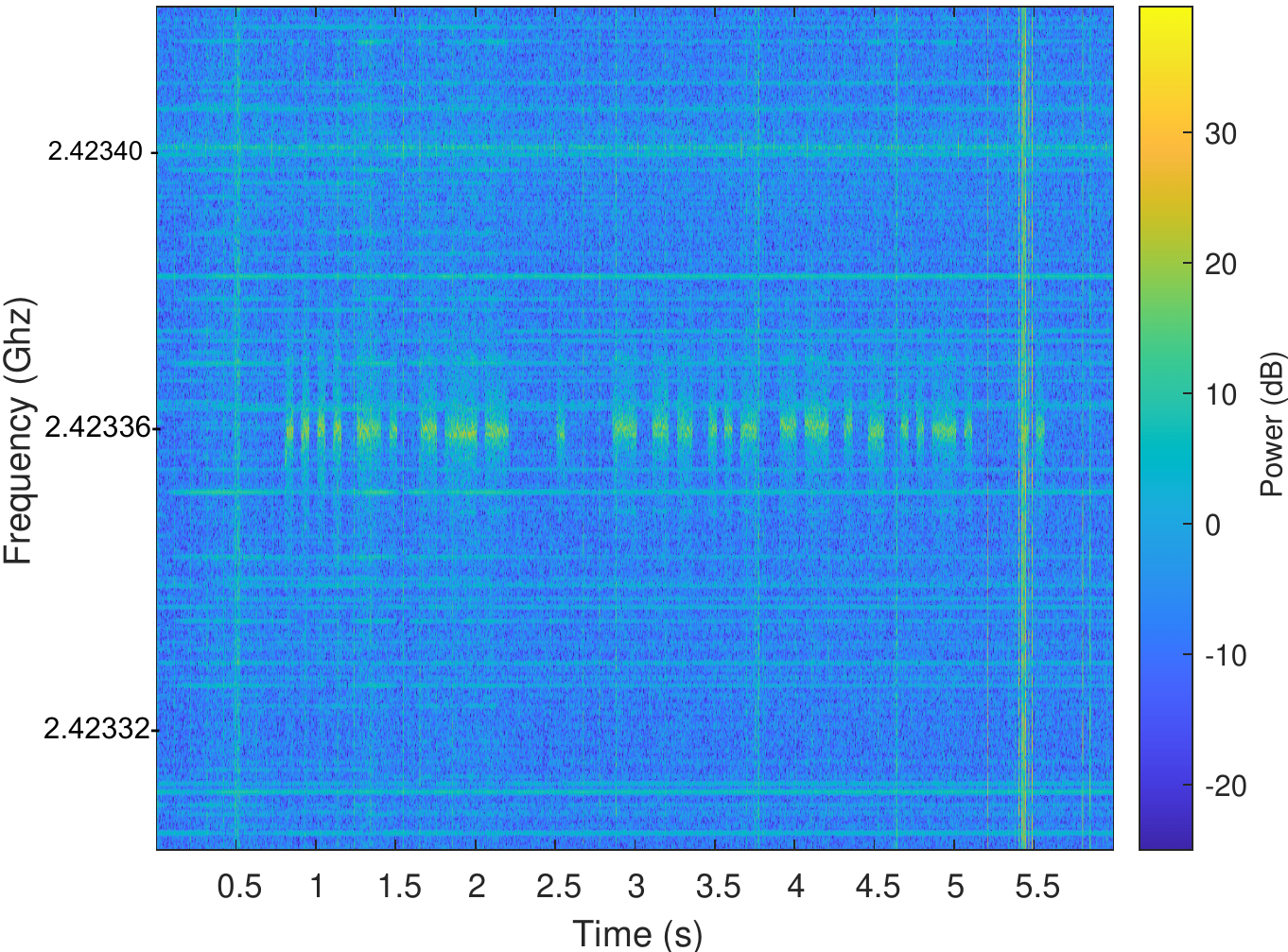}
	\caption{AIR-FI packet as transmitted from a workstation with a DDR4 (2400 MHz) memory module. The transmission overlaps channels 3,4, and 5.}
	\label{fig:message}
\end{figure}

Figure \ref{fig:message} shows an AIR-FI packet transmitted from a workstation with a DDR4 (2400 MHz) memory module. In this case, the transmission around ~2.42 GHz overlaps Wi-Fi channels 3,4, and 5. 

%
%
%

\section{Reception}
\label{sec:rec}
\begin{figure}  
	\centering
	\includegraphics[width=0.7\linewidth]{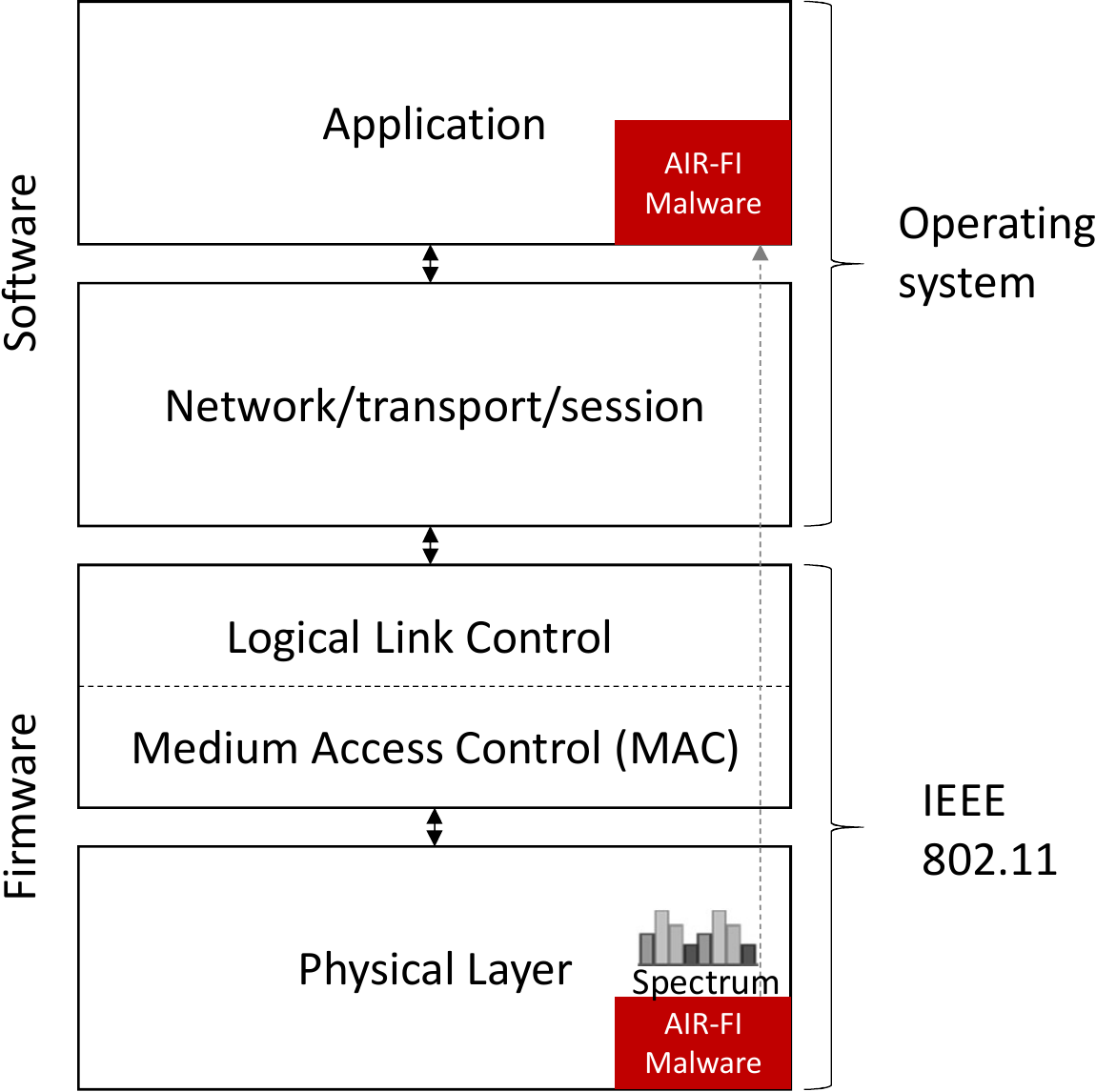}
	\caption{The receiver side of AIR-FI malware.}
	\label{fig:firmware}
\end{figure}

As shown in Section \ref{sec:trans}, the electromagnetic emissions generated by the data bus are around the 2.4 GHz frequency range and overlap the Wi-Fi channels. 
In Wi-Fi transceiver chips, the baseband processor handles the radio, PHY and MAC layers. The Internet, transport, and application layers are processed by the software protocol stack, usually in the kernel drivers. In order to measure the interference generated, the attacker has to access the low-level radio measurement information from the PHY layer.
This can be done by compromising the firmware of the Wi-Fi chips and passing the required radio measurements to the software stack. The architecture of AIR-FI malware is illustrated in Figure \ref{fig:firmware}. The firmware level code invokes the radio frequency (RF) information which is usually maintained through the Rx chain to reach the baseband processing. The data is passed to the AIR-FI at the application layer through operating system (e.g., via kernel module).     

\subsection{Wi-Fi Chip PHY Layer}
To access the radio and PHY layer data, we used the spectral analysis feature within Atheros 802.11n Wi-Fi chipsets. The Atheros chips (AR92xx and AR93xx) can report the data of the raw FFT measurement data from the baseband processor to the software stack. The data consists of vector of FFT bins for 56 subcarriers of the 20 MHz bandwidth channels. The data includes the absolute magnitude ($abs(i)+abs(q)$) for each bin, an index for the the strongest FFT bin, and the maximum signal magnitude.

\begin{figure}
	\centering
	\includegraphics[width=0.8\linewidth]{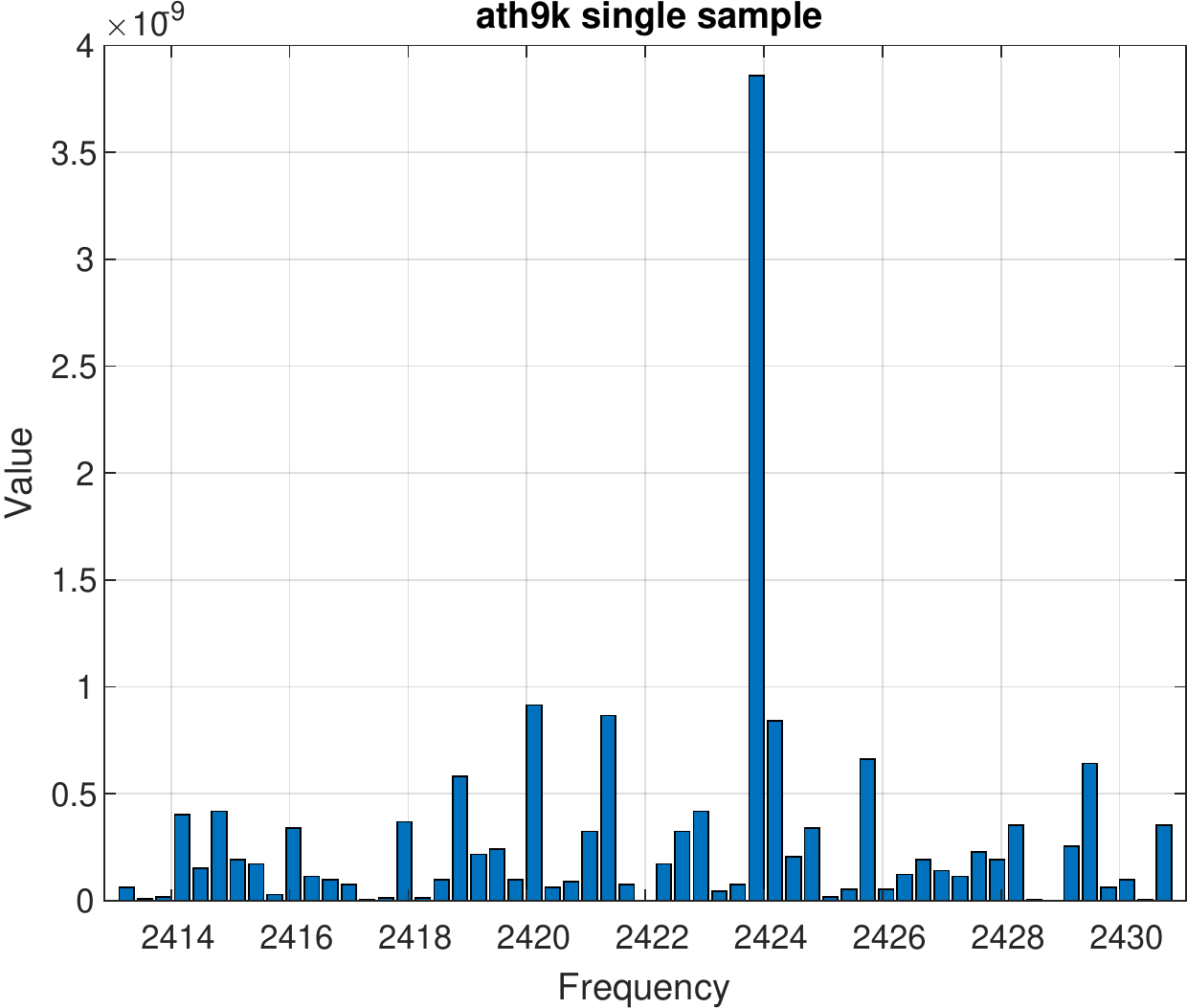}
	\caption{The FFT measurements of  Wi-Fi channel 3 as measured by the Atheros Wi-Fi receiver, with a transmission from the air-gapped computer. The signal can be seen in the 2424 MHz bin.}
	\label{fig:bins2}
\end{figure}

\begin{figure}
	\centering
	\includegraphics[width=0.8\linewidth]{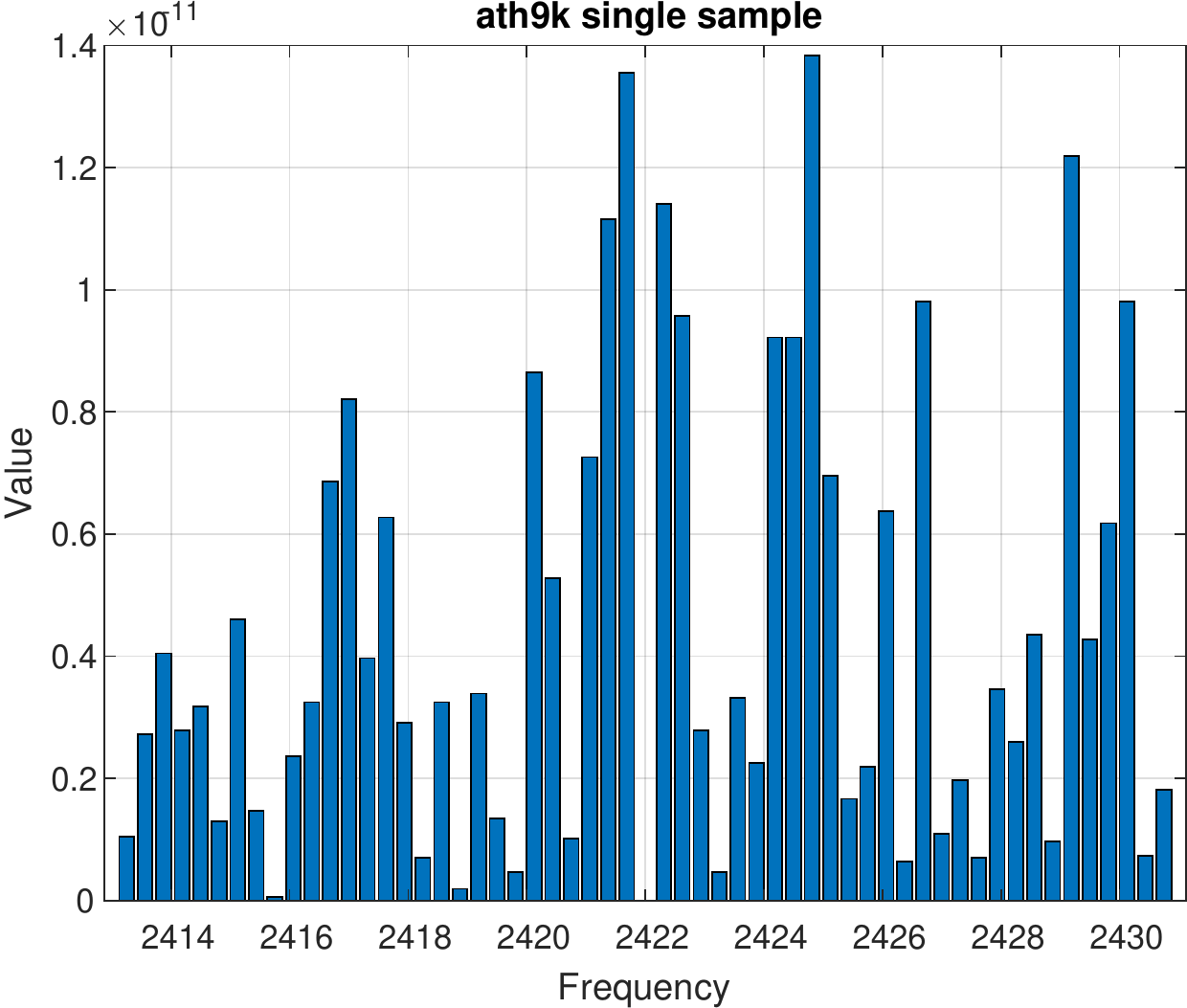}
	\caption{The FFT measurements of  Wi-Fi channel 3 as measured by the Atheros Wi-Fi receiver, without a transmission from the air-gapped computer.}
	\label{fig:bins1}
\end{figure}

Figures \ref{fig:bins2} and \ref{fig:bins1} show Wi-Fi channel 3 with and without AIR-FI transmission, respectively. The 56 bins of FFT are measured by the Atheros Wi-Fi chipset and delivered to the application layer. As can be seen, with the AIR-FI transmission, the amount of energy in the 2.424 GHz frequency bin is significantly higher than other bins in this channel with an SNR value of 9 dB.

\subsection{Reception Modes}
The Atheros chips support two main modes of reception: (1) scanning mode, and (2) triggering mode.

\subsubsection{Scanning mode}
In this mode the FFT information is returned for every Wi-Fi channel when a channel scan is performed. This can stop the Wi-Fi reception for the several hundred milliseconds it takes to scan the whole spectrum. This mode is maintained by setting the \texttt{chanscan} value to the \texttt{spectral\_scan\_ctl} control device. This mode can be used by the attacker to search for a covert transmission if the channel is unknown in advance.  

\subsubsection{Triggering mode}
In this mode, the FFT information is returned for a specific Wi-Fi channel when the Wi-Fi is operating. This mode is maintained by setting the value of the \texttt{spectral\_scan\_ctl} control device to \texttt{manual} and then initiating \texttt{trigger} commands. The scan samples are returned continuously from the channel currently configured. 

\begin{figure}
	\centering
	\includegraphics[width=0.8\linewidth]{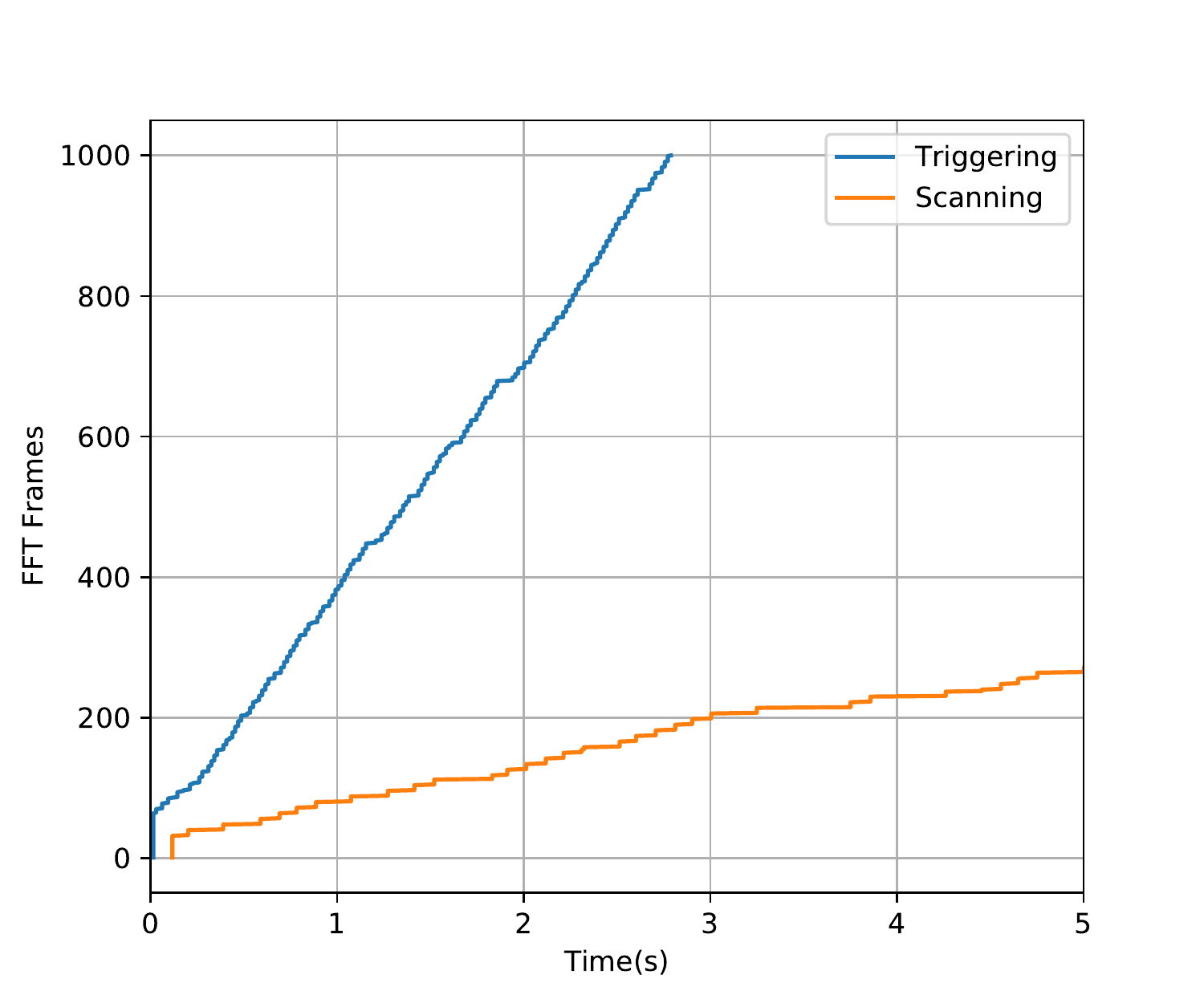}
	\caption{The number of FFT frames received in the scanning and triggering modes.}
	\label{fig:scantrig}
\end{figure}

\begin{figure}
	\centering
	\includegraphics[width=0.7\linewidth]{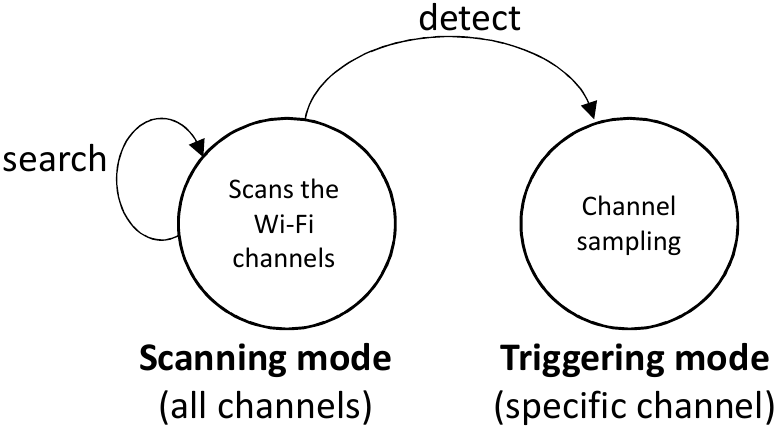}
	\caption{The transition between the scanning and triggering mode.}
	\label{fig:modes}
\end{figure}

As seen in Figure \ref{fig:scantrig}, the triggering mode is considerably faster than the scanning mode. The graph shows the number of FFT frames received in the scanning and triggering modes over a period of five seconds. The scanning mode can be used by malware to search the AIR-FI transmissions if the operational frequency is unknown in advance. After detecting a transmission, the malware can begin to operate in the triggering mode to receive the actual data (Figure \ref{fig:modes}). 

%
%

\subsection{Demodulation}
The pseudo code of the demodulator is presented in Algorithm \ref{algdemod}. We provide the implementation for a software defined radio (SDR) receiver. 

\paragraph{Atheros Wi-Fi Chip} Note that the implementation for the Atheros Wi-Fi receiver is based on the same concepts of the SDR code shown in Algorithm \ref{algdemod}. However, the Atheros implementation includes the extra steps of triggering the \texttt{spectral\_scan\_ctl} device, and receiving, buffing, decoding and parsing the FFT frames exposed by the Atheros chip. To simplicity the discussion and since we are not considering this as the main contribution of our work, we omitted the chip-specific details from our discussion.

\begin{algorithm}
	\caption{demodulate(deviceAddress, freq, sampleRate, bufferSize, bitTime, windowSize)} 
	\label{algdemod} 
	\begin{algorithmic}[1] 
		\State $ enabled \gets False $
		\State $ ctx \gets setupContext(deviceAddress) $
		\State $ rxbuf \gets setupRxBuf(ctx, freq, sampleRate, \gets  .., bufferSize) $
		\\
		\While { $ True $ }
		
		\State $ rxbuf.refill() $ 
		\State $ buffer = rxbuf.read() $
		
		\State $ windows = splitToWindows(buffer, windowSize) $
		
		\For {$ window\ in\ windows $}
		\State $ spectrum = welch(window)$
		\State $ sampleValue \gets spectrum[0] $
		\State $ sample \gets [getCurrentTime(), sampleValue] $
		\State $ samples.append(sample) $
		\EndFor
		\If {$not\ enabled$}
		\State $thresh, enabled \gets detectEnable(samples, bitTime)$
		\EndIf
		
		\While {$ enabled\ and\ enoughSamplesForBit  \gets (samples, bitTime) $}
		\State $ bit \gets samplesToBit(samples, bitTime, thresh) $
		\State $ output(bit) $
		\EndWhile
		
		\EndWhile
		
	\end{algorithmic}
\end{algorithm}

\begin{algorithm}
	\caption{detectEnable(samples, bitTime)} 
	\label{algdetect} 
	\begin{algorithmic}[1] 
		\State $ enableSequence \gets [1,0,1,0,1,0,1,0] $
		\State $ samplesDuration \gets getSamplesDuration(samples)$
		\State $ bitsInSamples \gets samplesDuration / bitTime$
		\If { $ bitsInSamples<2*len(enableSequence) $ }
		\State $ return\ 0, False $
		\EndIf
		\\
		
		\State $ calculatedCorr \gets calculateSampleCorrelationToBits  \gets (samples, enableSequence, bitTime) $
		\If { $ calculatedCorr<CORR\_THRESH $ }
		\State $ samples[:] \gets samples[1:] $
		\State $ return\ 0, False $
		\EndIf
		\\
		\State $ maxCorr \gets calculatedCorr$
		\State $ maxCorrIndex \gets 0 $
		
		\For {$ index\ in\ range(1, len(samples))$ }
		\State $ calculatedCorr \gets calculateSampleCorrelationToBits(samples[index:], enableSequence, bitTime) $
		\If {$calculatedCorr>maxCorr$}
		\State $maxCorr \gets calculatedCorr$
		\State $maxCorrIndex \gets index$
		\EndIf
		\EndFor
		\\
		\State $ samples[:] = samples[maxCorrIndex:] $
		\State $ enableSamples = extractEnableSamples(samples) $
		\State $ thresh = calculateThresh(enableSamples) $
		\State $ return\ thresh, True $
		
	\end{algorithmic}
\end{algorithm}

The OOK demodualtor is based on sampling and processing the FFT information for the specific Wi-Fi channel. In lines 2-3, the SDR device is initialized and the receiving buffer is configured with the frequency (in MHz) of the channel to monitor, the sampling rate, and the buffer size. The demodulator continuously samples the data in the required frequency and splits it into windows of \texttt{windowSize} size. For each window, the algorithm estimates the power spectral density using Welch's method (lines 9-14). It then detects the \textit{enable sequence} (\texttt{10101010}) using the \texttt{detectEnable} routine (Algorithm \ref{algdetect}), and determines the thresholds (amplitudes) for '1' and '0' bits (lines 15-18). Finally, the bits are demodulated and added to the output vector (lines 18-21).    

\section{Evaluation}
\label{sec:eval}
In this section, we present the analysis and evaluation of the AIR-FI covert channel. We describe the experimental setup, and test the different reception modes used to maintain the covert channel. We also evaluate the efficacy of the covert channel in virtualized environments.   

\begin{table*}[]
	\centering
	\caption{Receivers used in the evaluation}
	\label{tab:Receivers}
	\begin{tabular}{@{}lll@{}}
		\toprule
		Receiver \#   & Device          & Specs                                                                   \\ \midrule
		SDR & ADALM-PLUTO     & Frequency range from 325 MHz to 3.8 GHz, based on AD9363 transceiver                    \\
		Wi-Fi & TL-WN722N V1.10 & Frequency range from 2.4 GHz to 2.4835 GHz, 4 dBi detachable omni directional antenna \\ \bottomrule
	\end{tabular}
\end{table*}

\subsection{Experimental Setup}

\subsubsection{Receivers}
We used two types of receivers for  reception:
 
\begin{itemize}
	\item A software-defined radio (SDR) receiver. 
	\item A USB Wi-Fi network adapter. 
\end{itemize}

Table \ref{tab:Receivers} contains the specs of the receiver devices.  
The ADALM-PLUTO SDR is capable of sampling the Wi-Fi frequency band and has RF coverage from 325 MHz to 3.8 GHz. The TL-WN722N Wi-Fi USB wireless network adapter is equipped with the Atheros AR9271 chipset which supports spectral scan capabilities.
During the evaluation, we connected the receivers to a Lenovo ThinkCentre M93p workstation, with an Intel Core i7-4785T and Ubuntu 16.04.1 4.4.0 OS.

\subsubsection{Transmitters}
For the transmission we used the four types of off-the-shelf workstations listed in Table \ref{tab:setup1}. WORKSTATION-1 and WORKSTATION-2 were installed with two standard DDR4 2400 MHz modules. WORKSTATION-3 and WORKSTATION-4 were equipped with DDR3 modules (2133 MHz and 1600 MHz, respectively). WORKSTATION-3 and WORKSTATION-4 were used to evaluate the attack scenario in which the memory is maliciously overclocked to reach the Wi-Fi frequency band.

\begin{table*}[]
	\centering
	\caption{The workstations used for the evaluation}
	\label{tab:setup1}
	\begin{tabular}{@{}llll@{}}
		\toprule
		PC   & Hardware                                                                                                           & RAM                                                                                  & OS                                                                                               \\ \midrule
		
		WORKSTATION-1  & \begin{tabular}[c]{@{}l@{}}ASRock ATX DDR4 X99 Extreme4\\ CPU- Intel Core i7-6900K @ 3.2Ghz- 16 cores\end{tabular} & \begin{tabular}[c]{@{}l@{}}Crucial 4 * 4GB DDR4 SRAM\\ 2.4GHz RAM clock\end{tabular} & \begin{tabular}[c]{@{}l@{}}Ubuntu 18.04.1 \\ 4.15.0-72-generic\end{tabular}  \\
		
		WORKSTATION-2  & \begin{tabular}[c]{@{}l@{}}ASRock ATX DDR4 X99 Extreme4\\ CPU- Intel Core i7-6900K @ 3.2Ghz- 16 cores\end{tabular} & \begin{tabular}[c]{@{}l@{}} SK Hynix 4 * 4GB DDR4 SRAM\\ 2.4GHz RAM clock\end{tabular} & \begin{tabular}[c]{@{}l@{}}Ubuntu 18.04.1 \\ 4.15.0-72-generic\end{tabular}  \\
			
		WORKSTATION-3 (overclocked) & \begin{tabular}[c]{@{}l@{}}X99-UD4-CF\\ Intel Core i5-5820K\end{tabular}                                           & 4 * 4GB DIMM DDR3 2133MHz Micron                                                          & \begin{tabular}[c]{@{}l@{}}Ubuntu 18.04.2 \\ 5.0.0-36-generic\end{tabular}                       \\
		WORKSTATION-4 (overclocked) & \begin{tabular}[c]{@{}l@{}}H97M-D3H\\ Intel Core i7-4790\end{tabular}                                              & 4 * 4GB DIMM DDR3 1600MHz Hynix                                                      & \begin{tabular}[c]{@{}l@{}}Ubuntu 18.04.1 \\ 4.15.0-72-generic\end{tabular}                      \\
	                     \bottomrule
	\end{tabular}
\end{table*}

The following subsections present the results obtained for the four workstations. During the experiments we transmitted sequences of frame packets. We tested three receiver modes: (1) the SDR, (2) the Wi-Fi adapter operating in the scanning mode, and (3) the Wi-Fi adapter operating in the triggering mode. We measured the SNR values using the SDR receiver, and the BER values were measured using the SDR and Wi-Fi receiver.   
 

%
%
%

\subsection{WORKSTATION-1 (2.4 GHz)}
\begin{figure}
	\centering
	\includegraphics[width=0.8\linewidth]{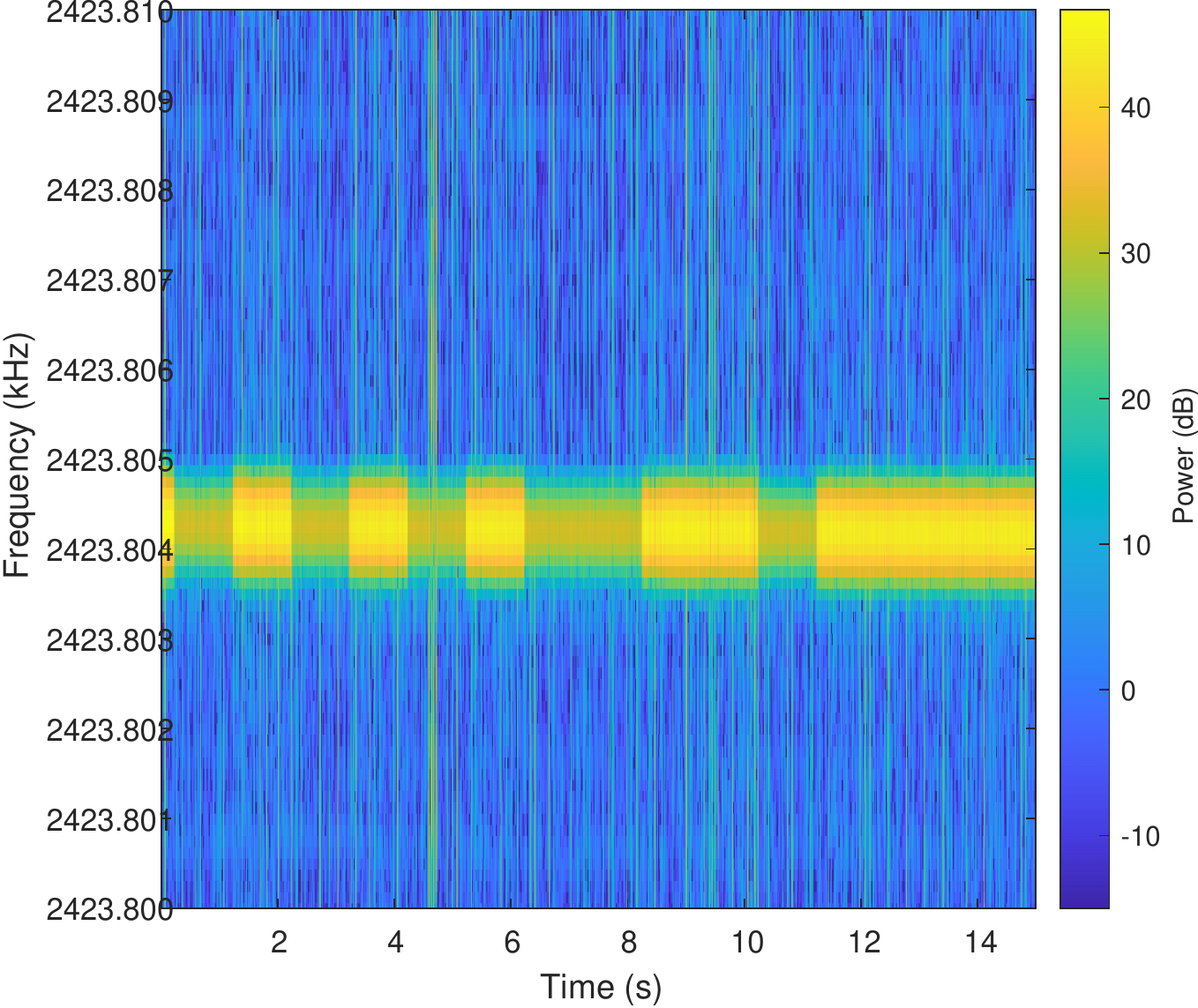}
	\caption{A transmission from WORKSTATION-1.}
	\label{fig:sig1}
\end{figure}

Figure \ref{fig:sig1} presents the signal generated from WORKSTATION-1 with all cores participating in the transmission. A signal with a bandwidth of 1 kHz exists in the  2423.804 - 2423.805 MHz range. In this case, the preamble sequence (10101010) can be seen at the beginning of the transmission. The signal generated by WORKSTATION-1 interferes with channels 3,4, and 5. 

\subsubsection{SDR}
Table \ref{tab:SNRBER-SDR-PC1} presents the 
signal-to-noise ratio (SNR) and bit error rate (BER) results, respectively, with WORKSTATION-1 as the transmitter and an SDR receiver as the receiver. In this case, we transmitted the data at a bit rate of 100 bit/sec and maintained a BER of 8.75\% for a distance up to 180 cm from the transmitter. Note that due to the local ramifications and interference, the signal quality may vary with the distance and location of the receiver.

\begin{table*}[]
	\centering
	\caption{The SNR and BER measured with WORKSTATION-1 (SDR receiver)}
	\label{tab:SNRBER-SDR-PC1}
	\begin{tabular}{@{}lllllllll@{}}
		\toprule
		Distance (cm) & 0    & 30   & 60   & 90     & 120    & 150  & 180  & 210    \\ \midrule
		SNR           & 14 dB & 10 dB & 13 dB & 5 dB    & 18 dB   & 13 dB & 20 dB & 3 dB    \\
		BER           & 0\%  & 0\%  & 0\%  & 0\% & 0\% & 0\%  & 8.75\%  & 22.5\% \\ \bottomrule
	\end{tabular}
\end{table*}

\subsubsection{Scanning \& triggering modes}
Table \ref{tab:BER-PC1}  presents the BER results with WORKSTATION-1 as the transmitter and the Wi-Fi dongle as the receiver when operating in the scanning and triggering modes. In scanning mode we were able to maintain BER of 0\% for the entire distance range of 0 - 180 cm, with a bit rate of 1 bit/sec. In the triggering mode we maintained a BER of 0 - 8.33\% for the range of 0 - 210 cm, with a bit rate of 10 bit/sec.

\begin{table*}[]
	\centering
	\caption{The BER measured with WORKSTATION-1 (Wi-Fi receiver)}
	\label{tab:BER-PC1}
	\begin{tabular}{lllllllllll}
		\hline
		Distance (cm) & 0   & 30     & 60     & 90  & 120    & 150 & 180 & 210  \\ \hline
		BER (scanning)    & 0\% & 0\%    & 0\%    & 0\% & 0\%    & 0\% & 0\% & 16.67\% \\ 
		BER (triggering)& 0\% & 0\%    & 0\%    & 8.33\% & 0\%    & 4.16\% & 0\% & 0\%  \\ \hline
	\end{tabular}
\end{table*}

\subsection{WORKSTATION-2 (2.4 GHz)}
\begin{figure}
	\centering
	\includegraphics[width=0.8\linewidth]{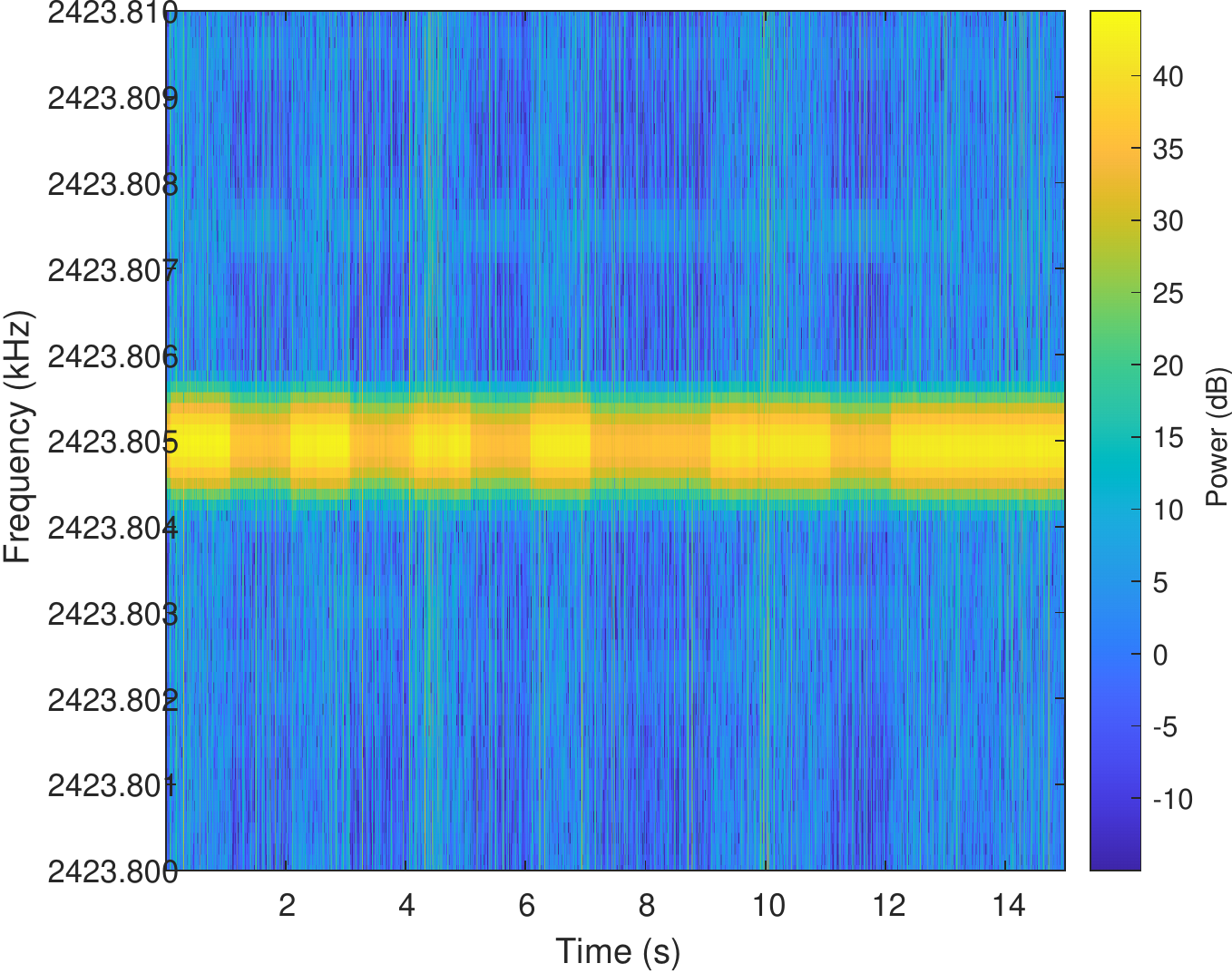}
	\caption{A transmission from WORKSTATION-2.}
	\label{fig:sig2}
\end{figure}

Figure \ref{fig:sig2} presents the signal generated from WORKSTATION-2  with all cores participating in the transmission. A signal with a bandwidth of 1 kHz exists in the 2423.8045 - 2423.8055 MHz frequency range. In this case, the preamble sequence (10101010) can be seen at the beginning of the transmission. The signal generated by WORKSTATION-2 overlaps with channels 3,4, and 5. 

\subsubsection{SDR}
Table \ref{tab:SNRBER-PC2} presents the SNR and BER results, respectively, with WORKSTATION-2 as the transmitter and an SDR receiver as the receiver. In this case, we transmitted the data at a bit rate of 100 bit/sec and were able to maintain a BER of 3.75\% for a distance up to 210 cm from the transmitter. Note that due to the local ramifications and interference, the signal quality might vary with the distance and location of the receiver.

\begin{table*}[]
	\centering
	\caption{The SNR and BER measured with WORKSTATION-2 (SDR receiver)}
	\label{tab:SNRBER-PC2}
	\begin{tabular}{@{}lllllllll@{}}
		\toprule
		Distance (cm) & 0    & 30   & 60   & 90     & 120    & 150  & 180  & 210    \\ \midrule
		SNR           & 13dB & 14dB & 6dB & 5dB    & 11dB   & 8dB & 10dB & 4dB    \\
		BER           & 0\%  & 0\%  & 0\%  & 1.25\% & 1.25\% & 0\%  & 0\%  & 3.75\% \\ \bottomrule
	\end{tabular}
\end{table*}

\subsubsection{Scanning \& triggering modes}
Table \ref{tab:BER-PC-2}  presents the BER results with WORKSTATION-2 as the transmitter and  a Wi-Fi dongle as the receiver in the scanning and triggering modes. In the scanning mode, we were able to maintain a BER of 0\% for the entire range of 0 - 270 cm, with a bit rate of 1 bit/sec. In the triggering mode, we were able to maintain a BER of 0 - 4.16\% for the range of 0 - 210 cm, with a bit rate of 10 bit/sec.

\begin{table*}[]
	\centering
	\caption{The BER measured with WORKSTATION-2 (Wi-Fi receiver)}
	\label{tab:BER-PC-2}
	\begin{tabular}{lllllllllll}
		\hline
		Distance (cm) & 0   & 30     & 60     & 90  & 120    & 150 & 180 & 210 & 240 & 270 \\ \hline
		BER (scanning)    & 0\% & 0\%    & 0\%    & 0\% & 0\%    & 0\% & 0\% & 0\% & 0\% & 0\% \\
		BER (triggering) & 0\% & 4.16\% & 4.16\% & 0\% & 4.16\% & 0\% & 0\% & 0\% & -   & -   \\ \hline
	\end{tabular}
\end{table*}

\subsection{WORKSTATION-3 (2133 MHz overclocked)}
The signal generated by WORKSTATION-3 resides in the 2402 MHz band which interferes with Wi-Fi channel 1.

Table \ref{tab:BER-PC3} IX presents the BER results with WORKSTATION-3 as the transmitter and the SDR receiver and Wi-Fi dongle as the receivers. In this case, a single core maintains the transmission. The workstation DRAM was overclocked to 2.4 GHz to target the Wi-Fi frequency bands. With the SDR receiver we were able to maintain a BER of 0\% for the entire range of 0 - 100 cm, with a bit rate of 100 bit/sec. With the Wi-Fi receiver in the scanning mode, we were able to maintain a BER of 0 - 0.15\% for the range of 0 - 100 cm, with a bit rate of 1 bit/sec. 

\begin{table*}[]
	\centering
	\caption{The BER measured with WORKSTATION-3 (SDR and Wi-Fi receivers)}
	\label{tab:BER-PC3}
	\begin{tabular}{llllllllllll}
		\hline
		Distance (cm)   & 0    & 10   & 20   & 30   & 40   & 50   & 60   & 70   & 80   & 90   & 100  \\ \hline
		BER (Pluto SDR) & 0\%    & 0\%    & 0.01\% & 0\%    & 0\%    & 0\%    & 0\%    & 0.02\% & 0\%    & 0.01\% & 0.17\% \\
		BER (Wi-Fi scanning)    & 0.01\% & 0.01\% & 0.02\% & 0.04\% & 0.04\% & 0.17\% & 0.25\% & 0.10\% & 0.11\% & 0.08\% & 0.15\% \\ \hline
	\end{tabular}
\end{table*}

Table \ref{tab:BER-triger-PC3} presents the BER results with WORKSTATION-3 as the transmitter and Wi-Fi dongle (triggering mode) as the receiver. In this case, we were able to maintain a BER of 0 - 14.7\% for the range of 0 - 300 cm, with a bit rate of 16 bit/sec.

\begin{table*}[]
	\centering
	\caption{The BER measured with WORKSTATION-3 (Wi-Fi receiver triggering)}
	\label{tab:BER-triger-PC3}
	\begin{tabular}{@{}llllllll@{}}
		\toprule
		Distance (cm) & 0   & 50   & 100  & 150 & 200 & 250  & 300  \\ \midrule
		BER (percent) & 6.8\% & 12.7\% & 14.7\% & 5.6\% & 4\%   & 11.9\% & 10.2\% \\ \bottomrule
	\end{tabular}
\end{table*}

\subsection{WORKSTATION-4 (1600 MHz overclocked)}
The signal generated by WORKSTATION-4 resides in the 2402 MHz band which interferes with Wi-Fi channel 1.

Table \ref{tab:BER-PC4} presents the BER results with WORKSTATION-4 as the transmitter and the SDR receiver and Wi-Fi dongle as the receivers. In this case, a single core maintains the transmission. With the SDR we were able to maintain a BER of 0\% for the entire range of 0 - 800 cm, with a bit rate of 100 bit/sec. With the Wi-Fi dongle, we were able to maintain a BER of 0 - 0.17\% for the range of 0 - 800 cm, with a bit rate of 1 bit/sec.
\begin{table*}[]
	\centering
	\caption{The BER measured with WORKSTATION-4 and SDR/Wi-Fi receivers}
	\label{tab:BER-PC4}
	\begin{tabular}{@{}llllllllll@{}}
		\toprule
		Distance (cm)   & 0    & 100  & 200  & 300  & 400  & 500 & 600 & 700  & 800 \\ \midrule
		BER (Pluto SDR) & 0    & 0    & 0    & 0.05 & 0    & 0   & 0   & 0    & 0   \\
		BER (Wi-Fi dongle)    & 0.04\% & 0.09\% & 0.02\% & 0.06\% & 0.17\% & 0\%   & 0.1\% & 0.08\% & 0\%   \\ \bottomrule
	\end{tabular}
\end{table*}

\subsection{Channels}
\begin{figure}
	\centering
	\includegraphics[width=1.0\columnwidth]{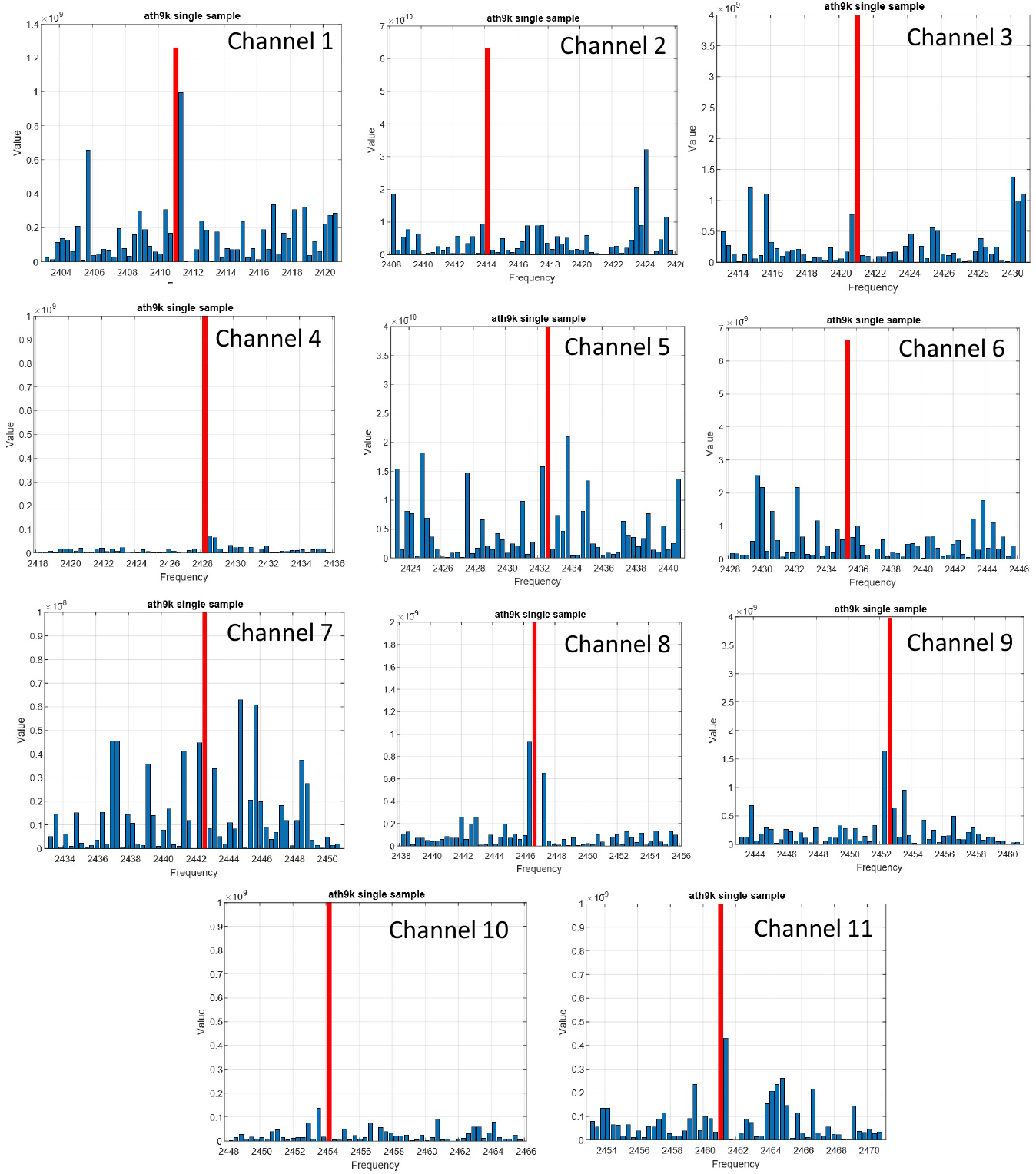}
	\caption{The FFT measurements of channels 1-11 as measured by the Atheros Wi-Fi receiver, with AIR-FI transmissions from WORKSTATION-1. The relevant bin is marked in red.}
	\label{fig:channels2}
\end{figure}

We measured the SNR values of AIR-FI transmission in 2.4 GHz Wi-Fi channels 1-11. 
Figure \ref{fig:channels2} shows the FFT measurements of channels 1-11 as measured by the Atheros Wi-Fi receiver, with a transmission from WORKSTATION-1. The AIR-FI signals can be seen in different frequencies of the channel. 
Table \ref{tab:csnr} summarizes the SNR values measured in each case. The SNR values ranged from 4.5 dB in channel 5 to 13 dB in channel 6.  

\begin{table*}[]
	\centering
	\caption{SNR values of AIR-FI transmission in channels 1-11}
	\label{tab:csnr}
	\begin{tabular}{@{}llllllllllll@{}}
		\toprule
		Channel   & 1         & 2         & 3         & 4         & 5         & 6         & 7         & 8         & 9         & 10        & 11        \\ \midrule
		AIR-FI frequency (GHz) & 2.411  & 2.414  & 2.421  & 2.428  & 2.432  & 2.436  & 2.442  & 2.446  & 2.452  & 2.454  & 2.461  \\
		SNR       & 5 dB      & 6 dB      & 11.5 dB   & 10 dB     & 4.5 dB    & 13 dB     & 8 dB      & 10 dB     & 8 dB      & 10 dB     & 10 dB     \\ \bottomrule
	\end{tabular}
\end{table*}

\subsection{Virtual Machines (VMs)}
\begin{figure}
 	\centering
 	\includegraphics[width=0.6\linewidth]{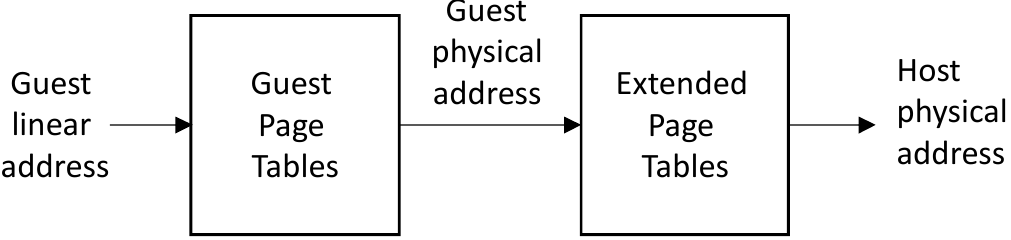}
 	\caption{The Extended Page Table (EPT) memory translation.}
 	\label{fig:EPT}
\end{figure}

\begin{figure}
	\centering
	\includegraphics[width=\linewidth]{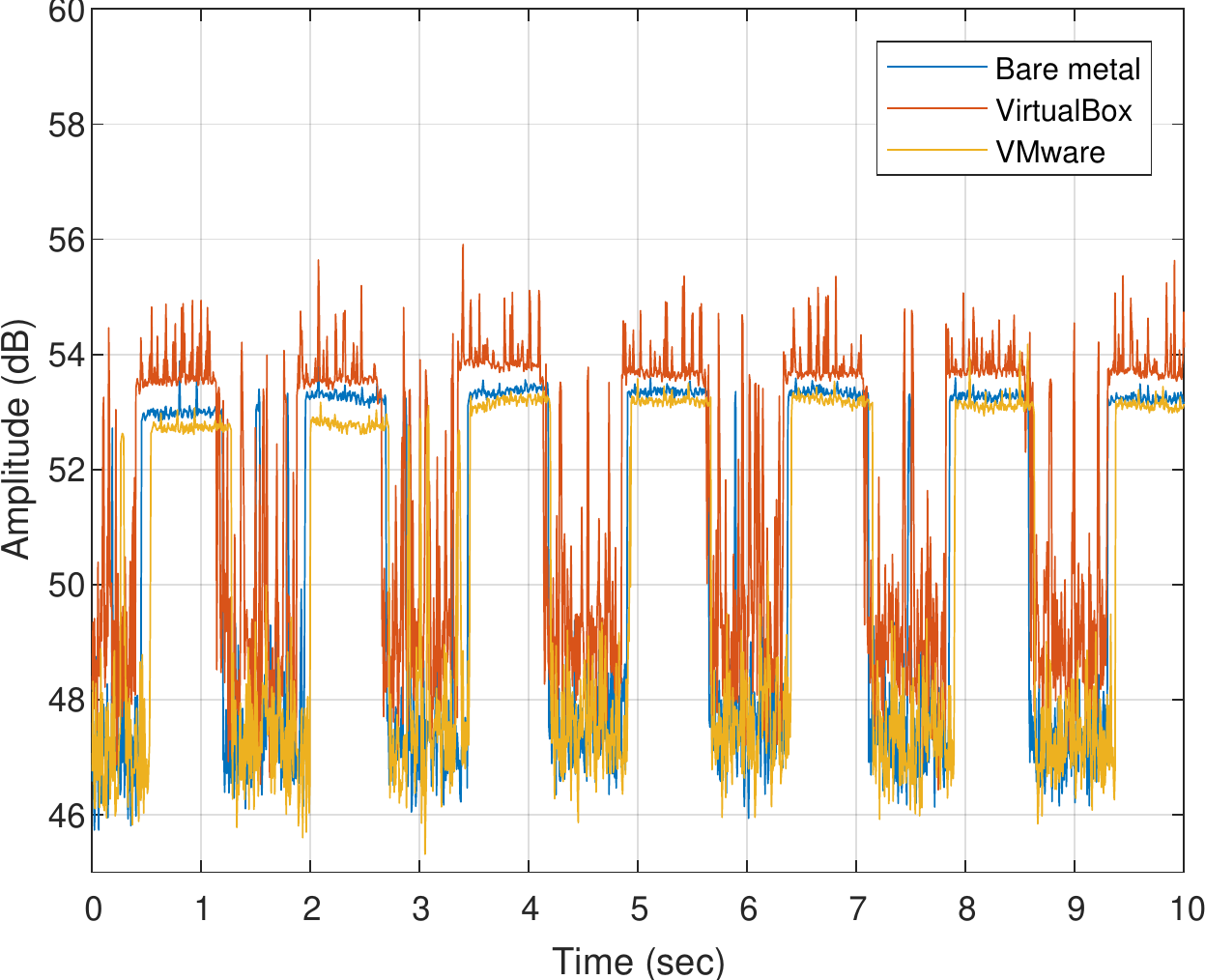}
	\caption{AIR-FI signal generated on bare metal, VMware and VirtualBox.}
	\label{fig:VMM}
\end{figure}

Virtualization technologies are commonly used in modern IT environments. One of their advantages is the isolation of hardware resources they enforce. Hypervisors/virtual machine monitors (VMMs) provide a layer of abstraction between the virtual machine and the physical hardware (CPU and peripherals). Since the covert channel is closely related to the memory access timing, we examined whether the virtualization layer caused interruptions and delays which may affect the signal quality. Generally speaking, in Intel VT-x, the mapping between the guest physical addresses and the host physical address is done through the extended page table (EPT). With the EPT, for each memory access operation, the MMU maps the guest linear address to the host physical address (Figure \ref{fig:EPT}). Note that the measurements show that this level of indirection may increase memory access latencies for some workloads [5]. We examined a transmission from WORKSTATION-1 and WORKSTATION-3 using three setups: a bare metal machine, a VMware VMM, and a VirtualBox VMM. Table XII contains details on the systems examined. Our experiments show that the covert signals can be maintained, even from within virtual machines. For WORKSTATION-1 and WORKSTATION-3, we measured a difference of at most 1 dB between the bare metal, VMWare, and VirtualBox signals (Figure \ref{fig:VMM}).

\begin{table*}[]
	\centering
	\caption{Virtualization}
	\label{tab:VMM}
	\begin{tabular}{@{}lllll@{}}
		\toprule
		\#         & Host                            & \multicolumn{1}{c}{VMM/Hypervisor}   & Guest                           & SNR (dB) \\ \midrule
		Bare metal & Ubuntu 18.04.1 5.3.0-53-generic & \multicolumn{1}{c}{N/A}              & \multicolumn{1}{c}{N/A}         & 5.09     \\
		Virtualbox & Ubuntu 18.04.1 5.3.0-53-generic & Virtualbox: 6.0.22 r137980           & Ubuntu 18.04.1 5.3.0-28-generic & 4.36     \\
		VMware     & Ubuntu 18.04.1 5.3.0-53-generic & VMware Player: 15.5.2 build-15785246 & Ubuntu 18.04.1 5.3.0-28-generic & 5.32     \\ \bottomrule
	\end{tabular}
\end{table*}

\section{Countermeasures}
\label{sec:counter}
There are several defensive approaches that can be used against the proposed covert channel.
 
\subsection{Separation}
The U.S and NATO telecommunication security standards (e.g., NSTISSAM TEMPEST/2-95 \cite{NSTISSAM75:online}) propose zone separation to protect against TEMPEST (Telecommunications Electronics Materials Protected from Emanating Spurious Transmissions) threats and other types of radiated energy attacks. In this approach, Wi-Fi transceivers are not allowed in certain classified areas. The NATO zoning procedure defines measures in which areas within a secured perimeter are classified as zone 0 to zone 3, depending on the safety requirements of the specific asset. In our case, Wi-Fi capable devices, such as smartphones, smartwatches, laptops, and so on, should be banned from the area of air-gapped systems.

\subsection{Runtime Detection} 
The signal generation algorithm is based on memory operations which trigger the DDR SDRAM emissions. Host based intrusion detection systems can monitor the activity of the processes in the OS. In our case, a process that abnormally performs memory transfer operations would be reported and inspected.  

A challenge to the runtime detection approach is that the signal generation algorithm (presented in Section \ref{sec:trans}) involves bare memory operations such as \texttt{memcpy()}. Monitoring the memory access instructions at runtime necessitates sandboxing or debugging of the process, which severely degrades performance \cite{guri2015gsmem} and can easily be bypassed by malware using rootkit techniques \cite{cardenas2011attacks}. In our case, the malware may inject a shellcode with a signal generation code into a legitimate, trusted process to bypass the security products. To overcome the evasion techniques, it is possible to employ solutions such as MemoryMonRWX, which is a bare metal hypervisor that can track and trap all types of memory access: read, write, and execute \cite{korkin2017detect}. However, all these detection techniques would likely suffer from high rates of false alarms, since many processes  intensively use the memory for legitimate needs (e.g., image processing, matrix calculations, etc.). 
Another approach is to use Wi-Fi monitoring hardware equipment in order to identify anomalies in the PHY layer of the Wi-Fi channels in the 802.11 bands \cite{WiFiOver87:online}. However, due to the legitimate activities of local access points and devices on the Wi-Fi channels such a detection approach will lead to many false positives. 

\subsection{Signal Jamming (hardware)} 
It is possible to block the covert channel by jamming the Wi-Fi frequency bands. Modern Wi-Fi jammers are signal blocking devices with radio frequency (RF) hardware which transmits radio waves in the entire range of Wi-Fi frequency bands (2.4 / 5 GHz). A typical Wi-Fi jammer generates high power, constant radio transmissions which span the channels and mask any legitimate Wi-Fi transmissions \cite{WiFiSign67:online}. 

\subsection {Signal Jamming (software)} 
\begin{figure}
	\centering
	\includegraphics[width=0.8\linewidth]{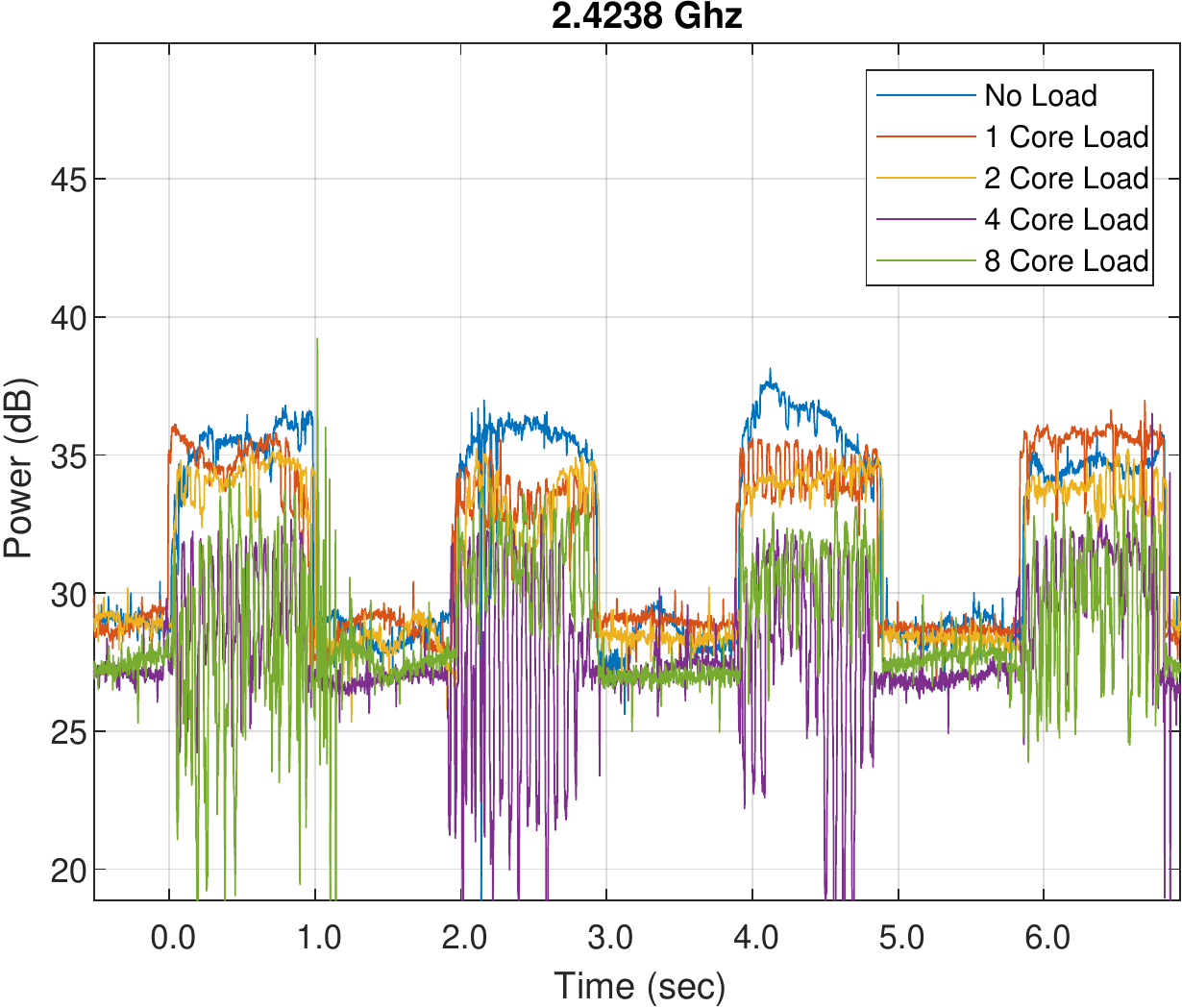}
	\caption{Signal jamming using intensive CPU operations.}
	\label{fig:jam2}
\end{figure}

In this approach, a background process which performs random memory or CPU operations is launched. The random workloads interfere with the execution of the malicious process and hence, interrupt the generation of the electromagnetic wave  emanated from the memory buses. Figure \ref {fig:jam2} shows the noise generated by WORKSTATION-1 when intensive prime number calculations were executed on one to eight cores using the \texttt{matho-primes} Linux command. Our measurements show that processes bound to six and eight cores, can significantly reduce the SNR of the original signal to SNR levels of 4.8 dB 3.1 dB, respectively. 


\subsection {Faraday Shielding} 
Faraday shielding is a special type of container used to block or limit the electromagnetic fields from interfering with or emanating from the shielded system.  
Faraday shielding copes with the threat presented in this paper by preventing
the leakage of Wi-Fi signals from the shielded
case. Generally, the computer shielding involves encompassing the
computer in a Faraday cage that does not permit stray
electromagnetic emanations. Physical isolation in which the whole room functions as an integral Faraday cage is also an option \cite{Faradayc67:online}. While this solution can be used in certain cases, it is impractical as a large-scale solution \cite{WiFiSign67:online}. 


\section{Conclusion}
\label{sec:conclusion}
In this paper, we demonstrated how attackers can exfiltrate data from air-gapped computers to a nearby Wi-Fi receiver via Wi-Fi signals. Our AIR-FI malware generates signals in the 2.4 GHz Wi-Fi frequency bands. The signals are generated through DDR SDRAM buses and do not require any special Wi-Fi hardware. Binary data can be modulated and encoded on top of the signals. We showed that a compromised nearby Wi-Fi device (e.g., smartphones, laptops, and IoT devices) can intercept these signals and decode the data. To extract the signals we utilized the low-level physical layer information that the Wi-Fi chips expose to the application layers. We implemented transmitters and receivers in different reception modes, and discussed design considerations and implementation details. We evaluated this covert channel in terms of bandwidth and distance and presented a set of countermeasures. Our results show that the covert channel can be effective at distances up to several meters from air-gapped computers. We achieved effective bit rates ranging from 1  to 100 bit/sec, depending on the type and mode of receiver used.

\Urlmuskip=0mu plus 1mu\relax
\balance
\bibliographystyle{IEEEtrans}
\bibliography{airfi,../AirGap,../AirGapCases,../mobile,../AirGapTools,../PowerSupply,../airgapsec,../attackvectors}

\end{document}